\documentclass[journal=jacsat,manuscript=article]{achemso}

\usepackage[version=3]{mhchem} 

\usepackage{amssymb}
\usepackage{bm}
\newcommand{\Z}{{\mathbb Z}}
\newcommand{\R}{{\mathbb R}}
\newcommand{\inner}[2]{{\left\langle{#1},{#2}\right\rangle}}
\renewcommand{\v}[1]{{\bm #1}}
\usepackage{soul}
\usepackage{color}

\author{Tomoyasu Yokoyama}
\affiliation{Technology Division, Panasonic Holdings Corporation, Osaka 571-8501, Japan}
\email{yokoyama.tomoyasu@jp.panasonic.com}
\author{Kazuhide Ichikawa }
\affiliation{Technology Division, Panasonic Holdings Corporation, Osaka 571-8501, Japan}
\author{Hisashi Naito}
\affiliation[Nagoya University]
{Graduate School of Mathematics, Nagoya University, Nagoya 464-8602, Japan}

\title[An \textsf{achemso} demo]
  {Crystal Structure Generation Based on Polyhedra using Dual Periodic Graphs}


\begin{document}






\begin{abstract}
Crystal structure design is important for the discovery of new highly functional materials because crystal structure strongly influences material properties.
Crystal structures are composed of space-filling polyhedra, which affect material properties such as ionic conductivity and dielectric constant.
However, most conventional methods of crystal structure prediction use random structure generation methods that do not take space-filling polyhedra into account, contributing to the inefficiency of materials development.
In this work, we propose a crystal structure generation method based on discrete geometric analysis of polyhedra information.
In our method, the shape and connectivity of a space-filling polyhedron are represented as a dual periodic graph, and the crystal structure is generated by the standard realization of this graph.
We demonstrate that this method can correctly generate face-centered cubic, hexagonal close-packed, and body-centered cubic structures from dual periodic graphs.
This work is a first step toward generating undiscovered crystal structures based on the target polyhedra, leading to major advances in materials design in areas including electronics and energy storage.
\end{abstract}

\section{Introduction}
The fundamental parameters that determine material properties are composition and structure. 
The smallest unit of composition is an atom, and the composition can be designed by changing the type and ratio of atoms. 
The smallest unit of structure can be regarded as a polyhedron, especially in crystal structures. 
This is because a crystal structure is an assembly of repeating units periodically in which each unit consists of atoms arranged at the vertices of a space-filling polyhedron.

The shape and connectivity of such polyhedra strongly affect material properties such as ionic conductivity and dielectric constant. 
For example, it has been shown theoretically that ionic crystals with a body-centered cubic (BCC) framework structure exhibit higher ionic conductivity than those with a face-centered cubic (FCC) or hexagonal close-packed (HCP) framework structure \cite{Wang2015}. 
Ionic conduction is caused by transitions of mobile ions through the central sites of the polyhedra in a crystal. 
The FCC structure is composed of tetrahedral and octahedral tiles. The coordination polyhedra affect the energies of ions at their central sites.
These different site energies lead to large energy gaps in the transition state for ion diffusion, which leads to low ionic conductivity.
The BCC structure is composed of tetrahedra only, and mobile ions can diffuse from one tetrahedral site to another without passing through octahedral sites.
Because the site energy remains constant, the energy barrier for mobile ion diffusion is lower in the BCC framework compared with that in the FCC  framework. 
Thus, structural design based on polyhedra is crucial for developing efficient materials.

Methods for designing crystal structures based on polyhedra have not yet been established. 
Most crystal structure prediction methods rely on random structure generation, in which atoms are randomly arranged.
We previously reported a method for predicting crystal structure by inputting the local structure as clusters and randomly arranging these clusters \cite{Yokoyama2021}; however, it is difficult to generate a structure with a targeted polyhedron. 
Recently, methods using generative models such as generative adversarial network (GAN), variational autoencoder (VAE), and diffusion models have been proposed to reverse engineer crystal structures from material properties \cite{Kim2020,Noh2019,Xie2021,Zeni2023}. 
However, this approach requires a huge amount of data to accurately reproduce a structure. 
Furthermore, it is difficult to predict extrapolation regions using such methods.

In this work, we use discrete geometric analysis called the theory of standard realization to construct a new method of crystal structure generation.
The theory of standard realization of topological crystals proposed by Kotani and Sunada \cite{Kotani2001} is based on the principle of least energy and graph theory, which gives the most symmetric crystal structure among all placements of given periodic graphs.
The standard realization satisfies the equilibrium condition and symmetry of each vertex.
The equilibrium condition is that the sum of the vectors of the edges starting from an edge is zero.
Symmetry is that each symmetry of the graph structure of a crystal lattice generates a Euclidean motion that preserves the crystal structure.
For example, graphene (periodic regular hexagonal lattices) and diamond structures are standard realizations of hexagonal and diamond lattices, respectively \cite{Naito2023}.
This method has yielded a new metallic carbon structure \cite{Itoh2009, Sunada2008} and 
a new carbon material structure with negative curvature \cite{tagami2014}. 
Although a wide variety of coordination numbers have been reported for crystalline materials, the application of this theory has so far been limited to three-coordinate carbon materials.

\begin{figure}[hbtp]
  \centering
  \includegraphics[width=1\textwidth]{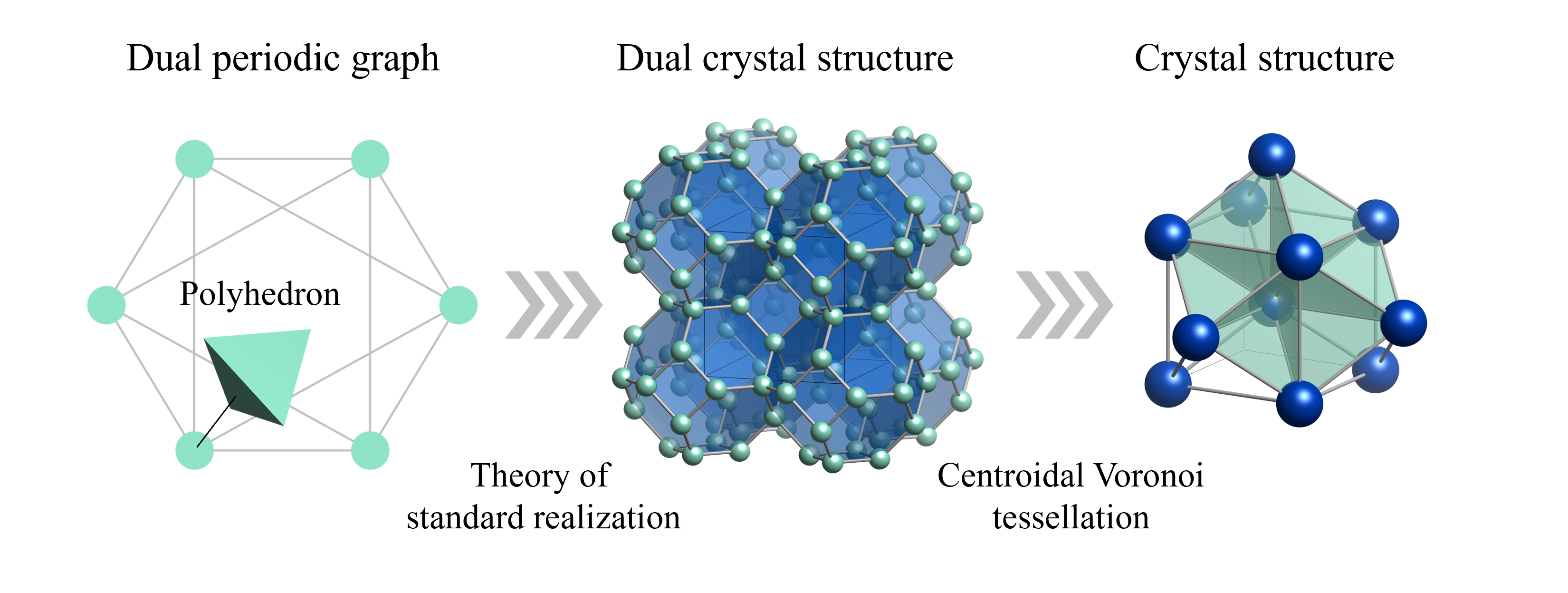}
  \caption{Conceptual diagram of our method for crystal structure generation. The dual periodic graph generated from the input polyhedra is transformed into a standard realization of a dual crystal structure. The obtained dual crystal structure is transformed into a crystal structure by centroidal Voronoi tessellation.}
  \label{fig:concept}
\end{figure}

This work aims to establish a new approach to generate crystal structures from polyhedron-based graphs using standard realization theory.
We use dual periodic graphs instead of periodic graphs to generate structures based on space-filling polyhedra.
A dual periodic graph, as we define it here, is a periodic graph obtained from a dual crystal structure.
A dual crystal structure is a structure formed by connecting the central points of space-filling polyhedra in a crystal structure.
The dual periodic graph directly represents the space-filling polyhedra in a crystal structure, which means that the dual periodic graph can be used to generate a crystal structure based on polyhedra.
A conceptual diagram of our proposed method is shown in Figure \ref{fig:concept}.
Given a dual periodic graph based on a space-filling polyhedron, we generate a dual crystal structure from the graph using standard realization theory and then transform the obtained dual crystal structure into a crystal structure.
Using our method, we demonstrate the generation of  FCC, HCP, and BCC structures, which are typical structures of metallic compounds.
A new finding in this work is the demonstration of the crystal structure generation from space-filling polyhedra by combining dual periodic graphs and the theory of standard realization.
We also discuss the challenges and potential of our method to discover new crystal structures based on polyhedra.

\section{Experimental Section}
\subsection{Standard Realizations}
A standard realization of a topological crystal, 
as proposed by Kotani and Sunada \cite{Kotani2001}, has the least energy obtained by the harmonic oscillator model, and is the most symmetric among all placements in the space for a 
given combinatorial structure.
Construction of a standard realization is a purely mathematical method.
In the following, the method to construct a standard realization is briefly described.
More detailed mathematical descriptions can be found elsewhere \cite{Kotani2001, Naito2009, Sunada2013, Naito2023}.
\par
Let $X_0 = (V_0, E_0 )$ be the dual graph of the target structure, 
and $X_T = (V_0, E_T)$ be a spanning tree of $X_0$, where $V_0$ is the set of vertices and $E_0$ are the set of edges.
The Betti number $b = b(X_0)$ is defined by $b = |E_0| - |E_T|$, 
and expresses the number of essential closed paths in $X_0$.
First, we set directions for all edges $\{e_j\}_{j=1}^{N}$ of $X_0$ ($N = |E_0|$), and
consider a space 
\begin{displaymath}
  C_1(X_0, \Z) = \{ a_1 e_1 + \cdots + a_N e_N \mid \{e_j\}_{j=1}^{N} = E_0, a_n \text{ are integers}\}. 
\end{displaymath} 
Then we may establish the notion of $\Z$-linear independence  for elements in $C_1(X_0, \Z)$.
A set of closed paths $\{\alpha_j\}_{j=1}^b$ in $X_0$, which are mutually $\Z$-linearly independent is selected, 
and then we can calculate the $N \times b$ matrix 
\begin{math}
  B = 
  \begin{bmatrix}
    \inner{e_j}{\alpha_k}
  \end{bmatrix}
\end{math}, 
the $b \times b$ matrix
\begin{math}
  G_0 = 
  \begin{bmatrix}
    \inner{\alpha_j}{\alpha_k}
  \end{bmatrix}
\end{math}, 
and
$A = G_0^{-1}B$, 
where $\inner{\cdot}{\cdot}$ is the inner product on $C_1(X_0, \Z)$, 
which is defined by $\inner{e_j}{e_k} = \delta_{jk}$ and $\inner{e_j}{\overline{e_k}} = -\delta_{jk}$.
Then, the matrix $A$ gives us the edges of a standard realization described by the basis  $\{\alpha_j\}_{j=1}^b$.
However, in general, a standard realization is constructed in the $b$-dimensional space $H_1(X_0, \R)$, 
which consists of linear combinations of all essential closed paths in $X_0$. Therefore, 
we should construct a projection onto the three-dimensional (3D) space.
\par
By selecting $\alpha_4, \cdots, \alpha_b$ as the basis of $(b-3)$-dimensional subspace $H$, 
defining the $3 \times 3$ matrix as 
\begin{displaymath}
  G = G_{11} - G_{12} G_{22}^{-1} G_{21}, 
  \quad
  G_0 = 
  \begin{bmatrix}
    G_{11} & G_{12}
    \\
    G_{21} & G_{22}
  \end{bmatrix}, 
\end{displaymath}
and projecting $A$ onto $H$, 
we obtain the 
edges of a standard realization in $3$D space described by the basis  $p_x$, $p_y$, $p_z$,  
where $G = \begin{bmatrix} \inner{\beta_j}{\beta_k} \end{bmatrix}$.
Next, vectors of parallel transformation $\v{p}_x$, $\v{p}_y$, and $\v{p}_z$ are calculated in $3$D Euclidean space; these vectors satisfy $G = \begin{bmatrix} \inner{\v{p}_j}{\v{p}_k} \end{bmatrix}$. This provides $3$D vectors that describe the edges of a standard realization. 
By fixing a vertex $\v{v}_0 = (0, 0, 0)$, and, 
for each vertex $v_k$, ($k = 1, \ldots, |V|-1$), 
taking the shortest path $c_k$ from $v_0$ to $v_k$ in the spanning tree $X_T$, 
we may calculate the coordinate $\v{v}_k$ along the path $c_k$.
Finally, we calculate coordinates of extra vertices using vectors of edges in $X_0\setminus X_T$.

A $b$-dimensional crystal lattice can be realized in 3D space by choosing a $(b-3)$-dimensional subspace $H$ from $H_1(X_0, \R)$ and by using the above method. However, in order to construct a good (symmetric) 3D crystal lattice, $H$ must be chosen appropriately, and for this purpose, it is important to choose the basis of $H_1(X_0, \R)$ appropriately.

\subsection{Dual Crystal Structure and Dual Periodic Graph}
\begin{figure}[hbtp]
  \centering
  \includegraphics[width=1\textwidth]{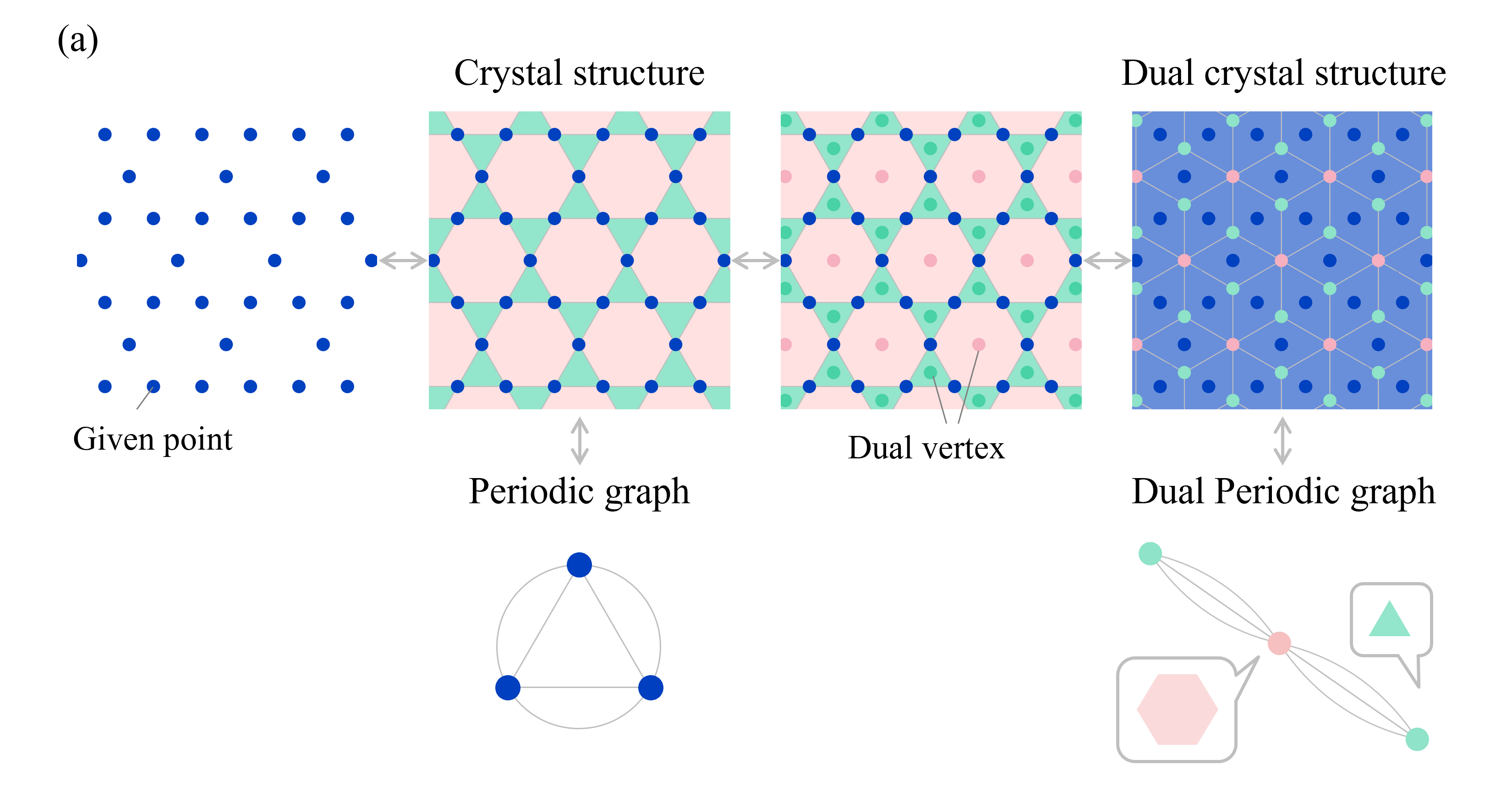}
  \includegraphics[width=1\textwidth]{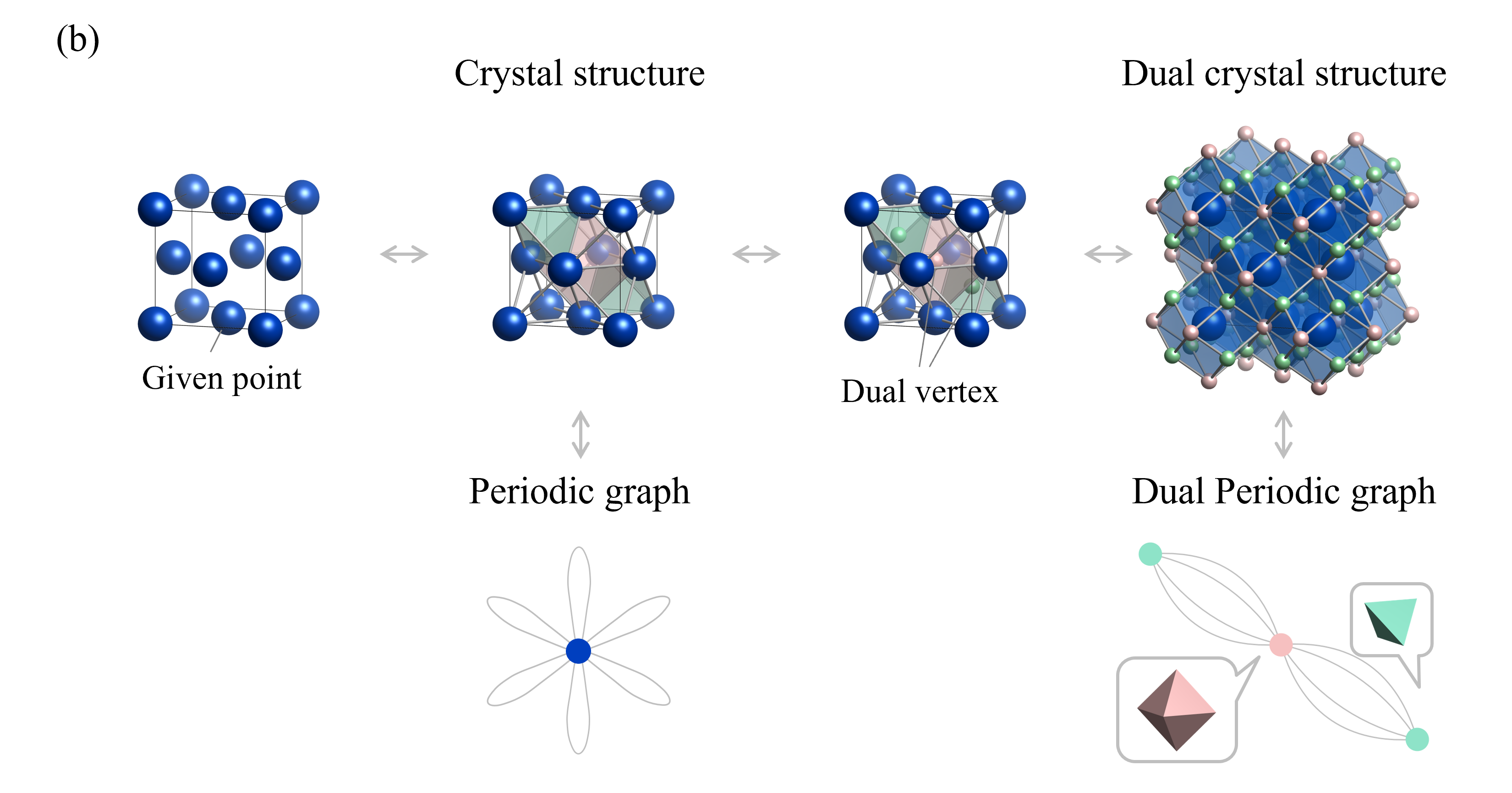}
  \caption{Relationship between crystal structures and dual crystal structures, periodic graphs, and dual periodic graphs in (a) two-dimensional planes and (b) three-dimensional space. The points in blue correspond to atomic sites, and the dual vertices shown in red and green correspond to interlattice sites. The crystal structure and dual crystal structure are obtained by connecting the given points and dual vertices, respectively. The periodic graph and dual periodic graph are obtained from the crystal structure and dual periodic graph, respectively.}
  \label{fig:dual}
\end{figure}

In general, a dual relationship is between two objects that are paired with each other; the dual relationship of a dual structure is the original structure.  
An octahedron and a cube or a Delaunay diagram and Voronoi diagram are in dual relationships.
An example of a two-dimensional ($2$D) crystal structure and its dual crystal structure is shown in Figure \ref{fig:dual}a.
This structure is the kagome lattice.
Note that a $b$-dimensional crystal structure here means a group of given points that repeat  periodically in $b$-dimensional space.
By connecting the given points with edges, we can see that this $2$D crystal structure is a tessellation with two triangles and one hexagon.
The periodic graph obtained from this crystal structure does not reveal that the $2$D crystal structure corresponding to this graph is composed of triangles and hexagons.

To explicitly consider these tiles in the graph, we introduce a dual crystal structure.
A dual crystal structure is a structure formed by connecting the central points of tiles in a crystal structure.
For example, the dual crystal structure of the kagome lattice is obtained by connecting the central points of the triangles and hexagons.
The dual periodic graph is the periodic graph obtained from the dual crystal structure.
The edges of the dual periodic graph correspond to the number of polygonal edges in the crystal structure. The dual periodic graph of the kagome lattice has three edges connecting to the green vertex and six edges connecting to the red vertex in Figure \ref{fig:dual}a, which correspond to the number of edges in the triangle and hexagon, respectively.
The symmetry of the original crystal structure and the dual crystal structure obtained by polyhedral packing are same. Therefore, the group of periodicity $X/X_0$ is also same for the structures constructed from each graph.

As in the case of the $2$D plane, dual crystal structures and dual periodic graphs can also be defined in $3$D space.
As an example, the $3$D crystal structure of the FCC lattice and its dual crystal structure are shown in Figure \ref{fig:dual}b.
Connecting the given vertices by edges, we can see that the primitive cell of this $3$D crystal structure consists of two tetrahedra and one octahedron.
The dual crystal structure of the FCC lattice is obtained by connecting the central points of the tetrahedra and octahedra.
Although the periodic graph of the FCC structure does not reveal that it is composed of two tetrahedra and an octahedron, the dual periodic graph has four edges connecting to the green vertex and eight edges connecting to the red vertex, which correspond to the face number of tetrahedra and octahedra, respectively.
Thus, a dual periodic graph represents a space-filling polyhedron, 
and a standard realization generates a dual crystal structure that satisfies the crystal structure of an appropriately selected space-filling polyhedra.

In general, the dual graph cannot be uniquely defined for a given graph, however, it can be uniquely defined by determining the placement of graph vertices in space. Our method defines bonds between atoms in a crystal. It then forms a dual crystal structure using these bonds and vertices. Finally, a dual periodic graph is obtained from that structure. In other words, the dual periodic graph is defined based on the polyhedral packing of the crystal structure. Our method uses centroidal  Voronoi tessellation (CVT) as this polyhedral packing. CVT is a method of defining a boundary between two adjacent points for a given point and decomposing the region by the boundary. 2D planes are decomposed into polygons, and 3D space is decomposed into polyhedra. If the intersection of such boundaries is defined as a dual vertex, CVT can transform from a crystal structure to a dual structure or from a dual structure to a crystal structure. If we can uniquely decompose a crystal structure into polyhedra as in the CVT, it is possible to uniquely define a dual graph. By using CVT as a polyhedral packing, a dual crystal structure with the same symmetry as the original crystal structure can be obtained for FCC, HCP, and BCC structures.

\subsection{Workflow}
\begin{figure}[hbtp]
  \centering
  \includegraphics[width=150pt]{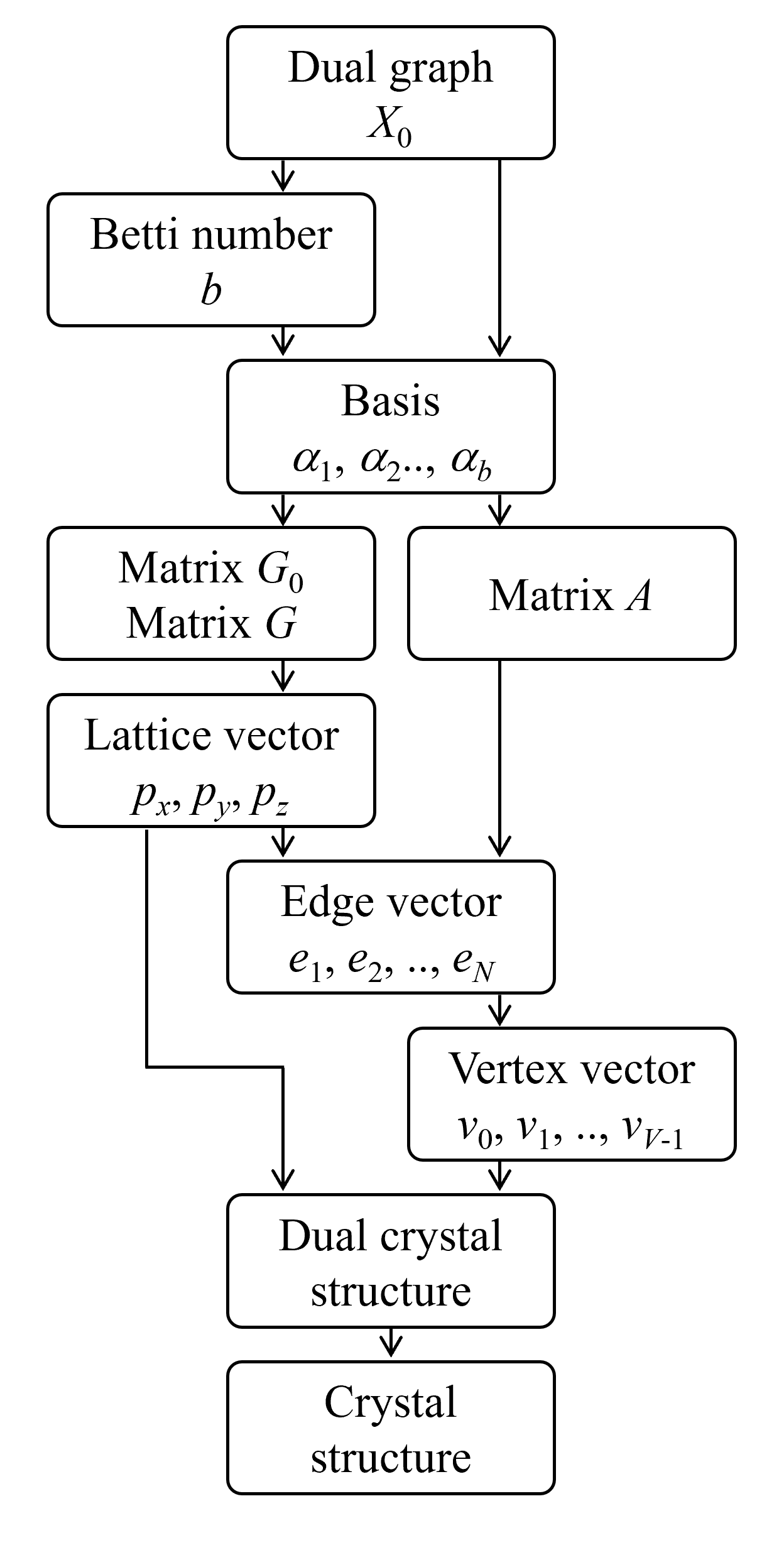}
  \caption{Flowchart of our method for crystal structure generation.}
  \label{fig:workflow}
\end{figure}

Figure \ref{fig:workflow} shows a flowchart of our method. 
In this method, a dual periodic graph is used as an input. The crystal structure corresponding to the dual periodic graph is generated in eight steps. Steps 1 to 7 describe the process of generating the dual structure from the dual periodic graph based on the theory of standard realization of topological crystals. Step 8 is the process of obtaining the crystal structure from the dual crystal structure. These steps are conducted as follows.

\begin{enumerate}
\item Calculate the Betti number $b$ from the given dual periodic graph $X_0$. Here, $b$ is equal to the number of edges in the graph minus the number of maximal spanning trees. In the case of the FCC dual graph, the number of edges is eight and the number of maximal spanning trees is two, so $b$ is six.
\item Define basis set $\{\alpha_j\}_{j=1}^b$ based on the number of selected closed paths corresponding to $b$ obtained in step 1.
\item Calculate the matrices $G_0$, $G$, and $A$ from the basis set $\{\alpha_j\}_{j=1}^b$ obtained in step 2 to project from the $b$-dimensional vector space to the $3$D vector subspace. The matrices $G_0$ and $G$ represent the conditions satisfied by the $b$- and $3$D lattice vectors, respectively. The matrix $A$ represents the basis of the edges in the $b$-dimensional vector space.
\item Define lattice vectors $\v{p}_x$, $\v{p}_y$, and $\v{p}_z$ such that the matrix $G$ obtained in step 3 is satisfied. The lattice vectors can also be obtained by Cholesky decomposition of the matrix $G$.
\item Calculate the edge vectors $\{e_j\}_{j=1}^N$ from the matrix $A$ obtained in step 3 and the lattice vectors $\v{p}_x$, $\v{p}_y$, and $\v{p}_z$ obtained in step 4.
\item Calculate the vertex vectors $\{v_k\}_{k=0}^{|V|-1}$ from the lattice vectors $\v{p}_x$, $\v{p}_y$, and $\v{p}_z$ obtained in step 4 and the edge vectors $\{e_j\}_{j=1}^N$ obtained in step 5. The vertex vectors correspond to the primitive coordinates.
\item Generate a dual crystal structure from the lattice vectors $\v{p}_x$, $\v{p}_y$, $\v{p}_z$ obtained in step 4 and the vertex vectors $\{v_k\}_{k=0}^{|V|-1}$ obtained in step 6.
\item Transform the dual structure obtained in step 7 into a crystal structure by CVT.
\end{enumerate}

The steps were performed using Python with NetworkX for graph operations \cite{Hagberg2008}, Numpy for operations of linear algebra \cite{Harris2020}, 
Pymatgen for generating the dual structure \cite{Ong2013}, and SciPy for computing CVT \cite{Virtanen2020}.
An example of Python code that generates a dual crystal structure from a dual periodic graph is shown in Supporting Information. 
Crystal structures were visualized by VESTA \cite{Momma2008}.

\section{Results and Discussion}
\subsection{Crystal Structure Generation for FCC Structures}
\begin{figure}[hbtp]
  \centering
  \includegraphics[width=1\textwidth]{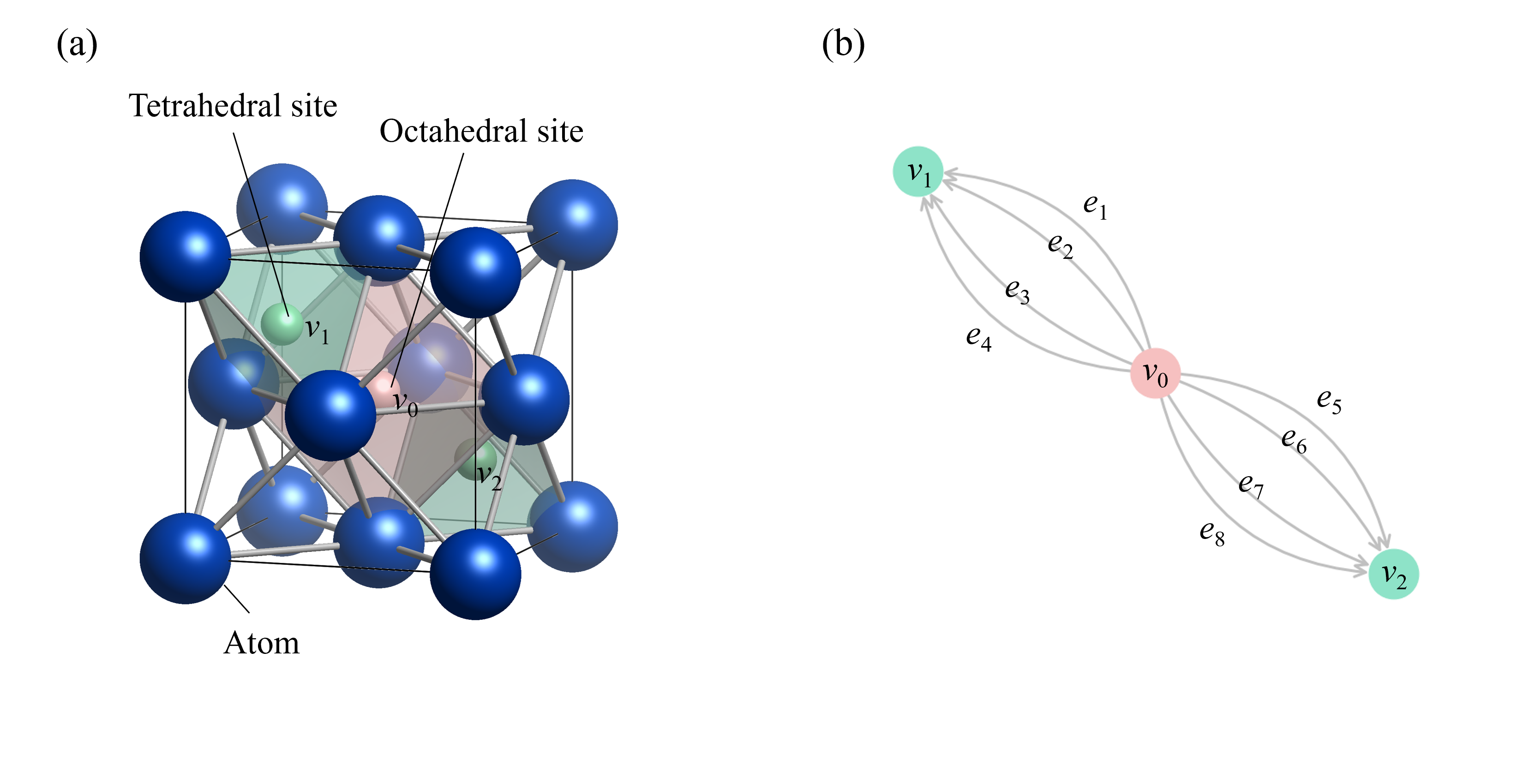}
  \includegraphics[width=1\textwidth]{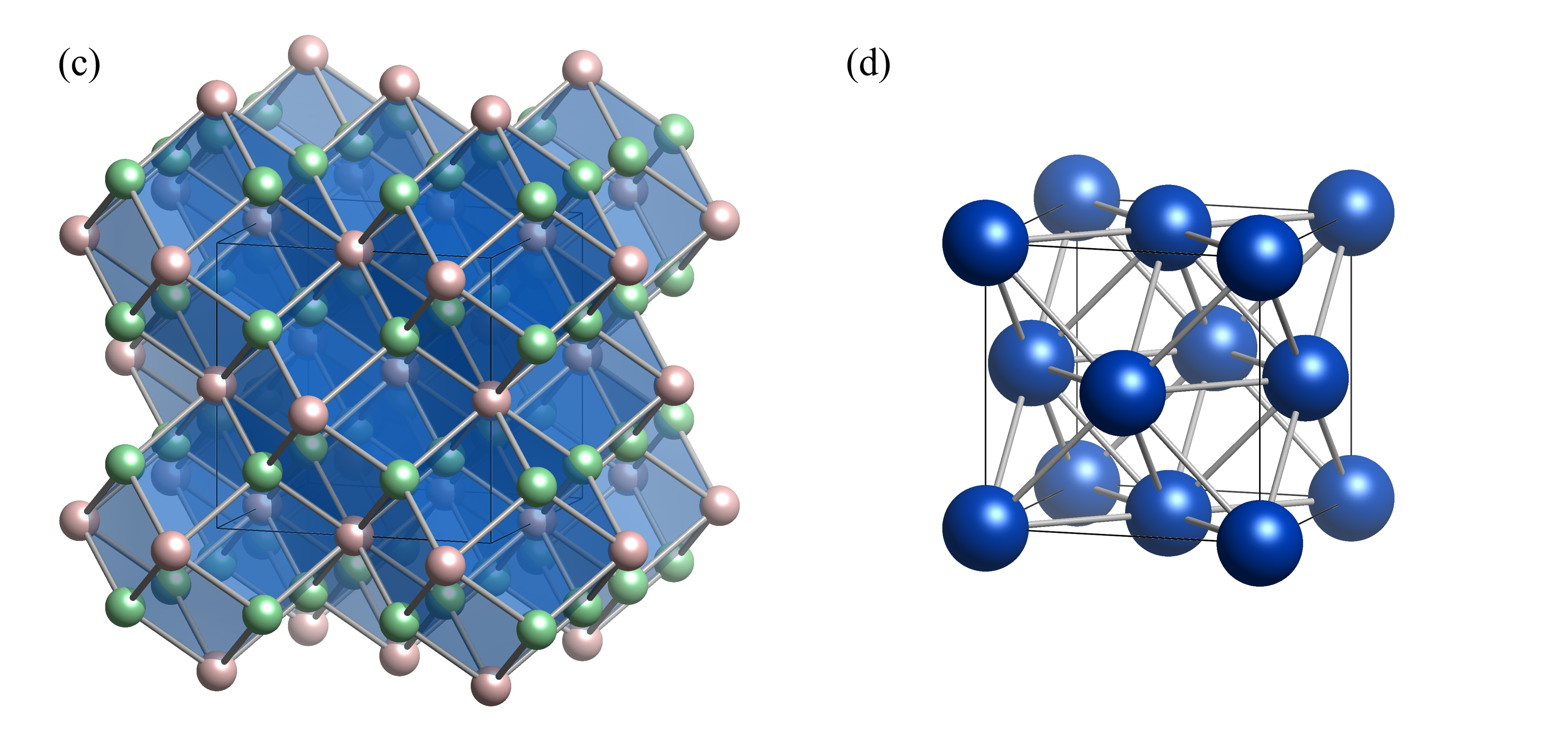}
  \caption{(a) The primitive cell of the FCC structure consists of one octahedron (red) and two tetrahedra (green). (b) Dual periodic graph of the FCC structure, where ${v_i}$ and ${e_i}$ are the vertex and edge numbers, respectively. (c) Dual crystal structure and (d) crystal structure generated from the dual periodic graph using our method.}
  \label{fig:fcc:structure}
\end{figure}

We demonstrated crystal structure generation from dual periodic graphs of FCC structures using our method. 
The FCC structure and dual periodic graph are shown in Figure \ref{fig:fcc:structure}a and b, respectively.
The primitive FCC structure has two interstitial tetrahedral sites and one interstitial octahedral site. 
In other words, the primitive FCC structure is tessellated by two tetrahedra and one octahedron. 
All the faces of each tetrahedron are shared with octahedra, and each octahedron shares all its faces with tetrahedra . 
The dual periodic graph has four edges connecting to the green vertex (${v_1}$ and ${v_2}$) and eight edges connecting to the red vertex (${v_0}$), which correspond to the face number of tetrahedra and octahedra, respectively. 

The specific process of generating the FCC structure from its dual periodic graph using standard realization theory is described below. For the FCC structure, the dual periodic graph $X_0$ has $b = 6$.
Taking a $Z$-basis of $H_1(X_0, \Z)$ as
\begin{displaymath}
  \begin{alignedat}{3}    
    {\alpha}_1 &= {e}_1-{e}_2, &\quad
    {\alpha}_2 &= {e}_1-{e}_3, &\quad
    {\alpha}_3 &= {e}_1-{e}_4, \\
    {\alpha}_4 &= {e}_1-{e}_2 + {e}_5-{e}_6, &\quad
    {\alpha}_5 &= {e}_1-{e}_3 + {e}_5-{e}_7, &\quad
    {\alpha}_6 &= {e}_1-{e}_4 + {e}_5-{e}_8, 
  \end{alignedat}
\end{displaymath}
The closed path corresponding to each basis is shown in Figure S1.
Then, by projecting onto the subspace spanning $\alpha_1$, $\alpha_2$, and $\alpha_3$, 
we obtain a standard realization in $3$D space  as 
\begin{equation}
  \label{eq:fcc:vector}
  \begin{alignedat}{3}
    e_1 &= \frac{1}{4}p_x + \frac{1}{4}p_y + \frac{1}{4}p_z, 
    &\quad
    e_5 &= -\frac{1}{4}p_x - \frac{1}{4}p_y - \frac{1}{4}p_z = - e_1, 
    \\
    e_2 &= -\frac{3}{4}p_x + \frac{1}{4}p_y + \frac{1}{4}p_z, 
    &\quad
    e_6 &= \frac{3}{4}p_x - \frac{1}{4}p_y - \frac{1}{4}p_z = - e_2,  
    \\
    e_3 &= \frac{1}{4}p_x - \frac{3}{4}p_y + \frac{1}{4}p_z, 
    &\quad
    e_7 &= -\frac{1}{4}p_x + \frac{3}{4}p_y - \frac{1}{4}p_z = - e_3,  
    \\
    e_4 &= \frac{1}{4}p_x + \frac{1}{4}p_y - \frac{3}{4}p_z,
    &\quad
    e_8 &= -\frac{1}{4}p_x - \frac{1}{4}p_y + \frac{3}{4}p_z = - e_4.
  \end{alignedat}
\end{equation}
The period vectors (a basis of Bravais lattice) $\{\v{p}_x, \v{p}_y, \v{p}_z\}$ satisfies 
\begin{equation}
  \label{eq:fcc:gram}
  G
  =
  \begin{bmatrix}
    |p_x|^2 & \inner{p_x}{p_y} & \inner{p_x}{p_z} \\
    \inner{p_y}{p_z} & |p_y|^2 & \inner{p_y}{p_z} \\
    \inner{p_z}{p_x} & \inner{p_z}{p_y} & |p_z|^2 \\
  \end{bmatrix}
  =
  \frac{1}{2}
  \begin{bmatrix}
    2 & 1 & 1 \\
    1 & 2 & 1 \\
    1 & 1 & 2 \\
  \end{bmatrix}.
\end{equation}
Take $\v{p}_x$, $\v{p}_y$, and $\v{p}_z$
\begin{equation}
  \label{eq:fcc:lattice}
  \v{p}_x
  =
  (1, 0, 1), 
  \quad
  \v{p}_y
  =
  (1, 1, 0), 
  \quad
  \v{p}_z
  =
  (0, 1, 1), 
\end{equation}
which satisfies Equation (\ref{eq:fcc:gram}). 
Substituting Equation (\ref{eq:fcc:lattice}) into Equation (\ref{eq:fcc:vector}), 
we obtain positions of vertices as
\begin{displaymath}
  \begin{alignedat}{4}
    \v{v}_0 &= \left(0, 0, 0\right), 
    \\
    \v{v}_1 &= \v{v}_0 + \v{e}_1 = \left(\frac{1}{2}, \frac{1}{2}, \frac{1}{2}\right), 
    &\quad
    \v{v}_2 &= \v{v}_0 + \v{e}_2 = \left(-\frac{1}{2}, \frac{1}{2}, -\frac{1}{2}\right), 
    \\
    \v{v}_3 &= \v{v}_0 + \v{e}_3 = \left(\frac{1}{2}, -\frac{1}{2}, -\frac{1}{2}\right), 
    &\quad
    \v{v}_4 &= \v{v}_0 + \v{e}_4 = \left(-\frac{1}{2}, -\frac{1}{2}, \frac{1}{2}\right), 
    \\
    \v{v}_5 &= \v{v}_0 + \v{e}_5 = \left(-\frac{1}{2}, -\frac{1}{2}, -\frac{1}{2}\right), 
    &\quad
    \v{v}_6 &= \v{v}_0 + \v{e}_6 = \left(\frac{1}{2}, -\frac{1}{2}, \frac{1}{2}\right), 
    \\
    \v{v}_7 &= \v{v}_0 + \v{e}_7 = \left(-\frac{1}{2}, \frac{1}{2}, \frac{1}{2}\right), 
    &\quad
    \v{v}_8 &= \v{v}_0 + \v{e}_8 = \left(\frac{1}{2}, \frac{1}{2}, -\frac{1}{2}\right).
  \end{alignedat}
\end{displaymath}

The dual crystal structure and crystal structure obtained from this graph are shown in Figure\ref{fig:fcc:structure}c and d, respectively. 
The obtained crystal structure is consistent with the FCC structure. 
Therefore, the FCC structure can be reproduced from the dual periodic graph using our method.
\subsection{Crystal Structure Generation for HCP Structures}
\begin{figure}[hbtp]
  \centering
  \includegraphics[width=1\textwidth]{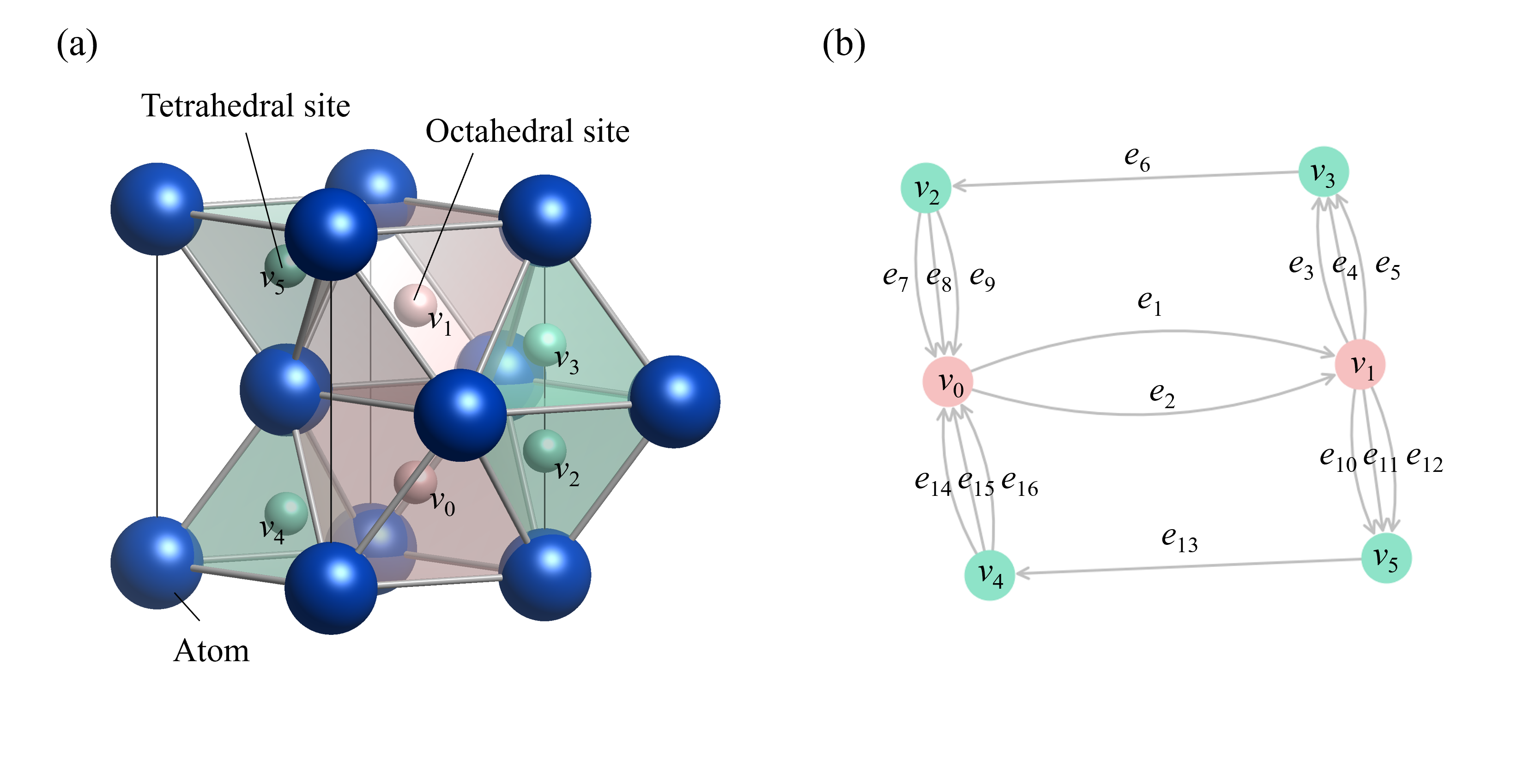}
  \includegraphics[width=1\textwidth]{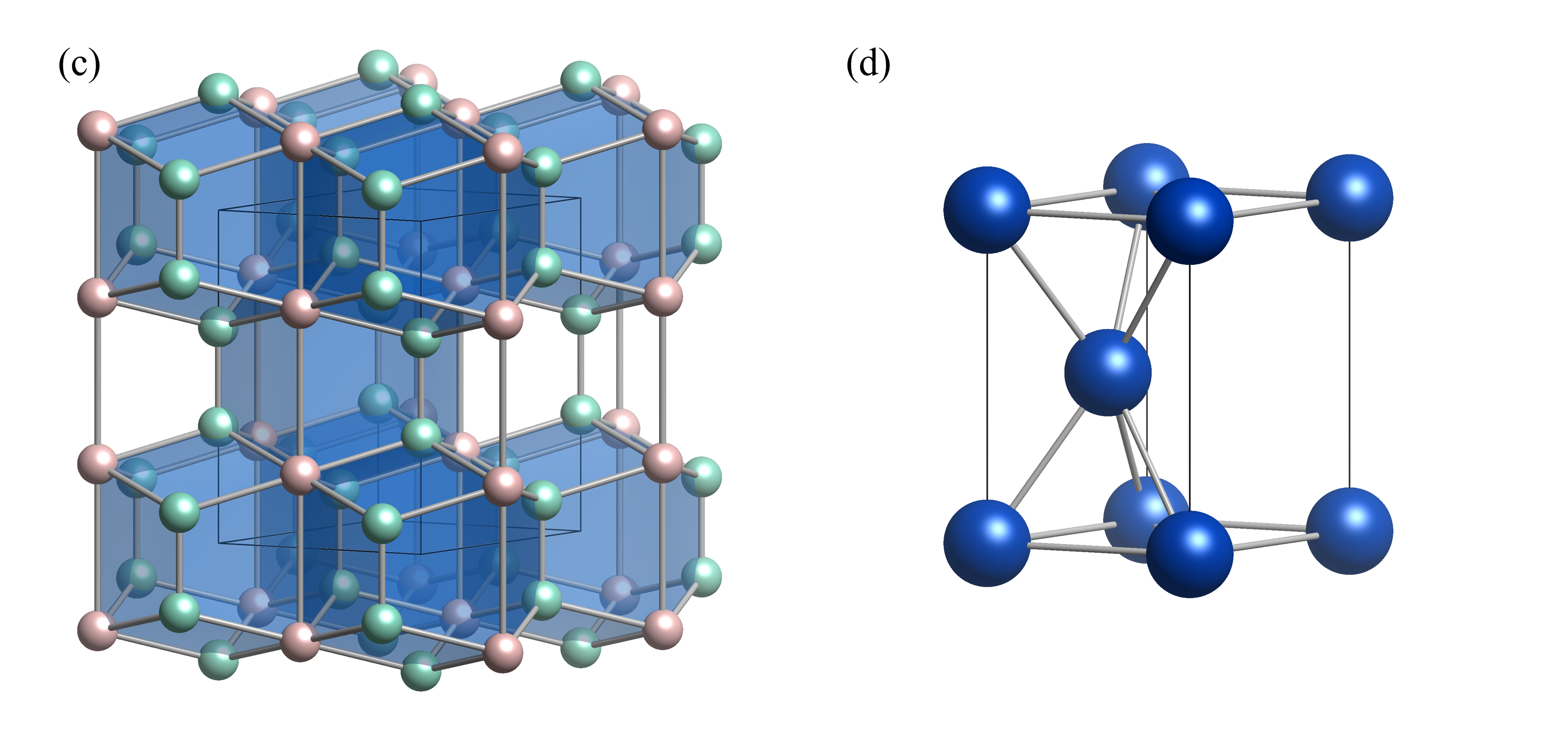}
  \caption{(a) The primitive cell of the HCP structure consisting of two octahedra (red) and four tetrahedra (green). (b) Dual periodic graph of the HCP structure, where ${v_i}$ and ${e_i}$ are the vertex and edge numbers, respectively. (c) Dual crystal structure and (d) crystal structure generated from the dual periodic graph using our method.}
  \label{fig:hcp:structure}
\end{figure}

We also demonstrated crystal structure generation from dual periodic graphs of HCP structures using our method. 
The HCP structure and dual periodic graph are shown in Figure \ref{fig:hcp:structure}a and b, respectively. 
The HCP structure has four interstitial tetrahedral sites and two interstitial octahedral sites. 
In other words, the primitive HCP structure is tessellated by four tetrahedra and two octahedra.
Unlike the FCC structure, tetrahedra share faces with each other, and octahedra share faces with each other in the HCP structure.
The dual periodic graph has four edges connecting to the green vertex (${v_2}$, ${v_3}$, ${v_4}$, and ${v_5}$) and eight edges connecting to the red vertex (${v_0}$ and ${v_1}$), which correspond to the face number of tetrahedra and octahedra, respectively. 

The specific process of generating the FCC structure from its dual periodic graph using standard realization theory is described below. The dual periodic graph $X_0$ of HCP structure had $b = 11$.
A $\Z$-basis of $H_1(X_0, \Z)$ was taken as
\begin{displaymath}
  \begin{alignedat}{3}    
    \alpha_1 &= e_{1} - e_{2}, 
    &\quad
    \alpha_2 &= e_{7} - e_{8}, 
    &\quad
    \alpha_3 &= e_{4} - e_{5}, 
    \\
    \alpha_4 &= e_{7} - e_{8} + e_{15} - e_{16}, 
    &\quad
    \alpha_5 &= e_{4} - e_{5} + e_{10} - e_{12},
    \\
    \alpha_6 &= e_{1} + e_{6} + e_{3} + e_{7},
    &\quad
    \alpha_7 &= e_{1} + e_{6} + e_{4} + e_{9},
    \\
    \alpha_8 &= e_{1} + e_{6} + e_{5} + e_{8},
    &\quad
    \alpha_9 &= e_{2} + e_{13} + e_{10} + e_{14},
    \\
    \alpha_{10} &= e_{2} + e_{13} + e_{11} + e_{15},
    &\quad
    \alpha_{11} &= e_{2} + e_{13} + e_{12} + e_{16}, 
  \end{alignedat}
\end{displaymath}
The closed path corresponding to each basis is shown in Figure S2.
Then, projecting onto the subspace spanning $\alpha_1$, $\alpha_2$, and $\alpha_3$, 
we obtain a standard realization in the $3$D space as 
\begin{equation}
  \label{eq:hcp:vector}
  \begin{alignedat}{3}
    e_1 &= \frac{1}{2}p_x, 
    &\quad
    e_2 &= -\frac{1}{2}p_x, \\
    e_3 &= -\frac{1}{10}p_x - \frac{2}{3} p_y - \frac{1}{3} p_z, 
    &\quad
    e_4 &= -\frac{1}{10}p_x + \frac{1}{3} p_y + \frac{2}{3} p_z, \\
    e_5 &= -\frac{1}{10}p_x - \frac{1}{3} p_y - \frac{1}{3} p_z, 
    &\quad
    e_6 &= -\frac{3}{10}p_x, \\
    e_7 &= -\frac{1}{10}p_x + \frac{2}{3} p_y + \frac{1}{3} p_z, 
    &\quad
    e_8 &= -\frac{1}{10}p_x + \frac{2}{3} p_y - \frac{1}{3} p_z, \\
    e_9 &= -\frac{1}{10}p_x - \frac{1}{3} p_y - \frac{2}{3} p_z, 
    &\quad
    e_{10} &= \frac{1}{10}p_x - \frac{1}{3} p_y - \frac{2}{3} p_z, \\
    e_{11} &= \frac{1}{10}p_x + \frac{2}{3} p_y + \frac{1}{3} p_z, 
    &\quad
    e_{12} &= \frac{1}{10}p_x - \frac{1}{3} p_y - \frac{1}{3} p_z, \\
    e_{13} &= \frac{3}{10}p_x, 
    &\quad
    e_{14} &= \frac{1}{10}p_x + \frac{1}{3} p_y + \frac{2}{3} p_z, \\
    e_{15} &= \frac{1}{10}p_x - \frac{2}{3} p_y - \frac{1}{3} p_z, 
    &\quad
    e_{16} &= \frac{1}{10}p_x + \frac{1}{3} p_y - \frac{1}{3} p_z, \\
  \end{alignedat}
\end{equation}
The period vectors (a basis of Bravais lattice) $\{\v{p}_x, \v{p}_y, \v{p}_z\}$ satisfies 
\begin{equation}
  \label{eq:hcp:gram}
  G
  =
  \begin{bmatrix}
    |p_x|^2 & \inner{p_x}{p_y} & \inner{p_x}{p_z} \\
    \inner{p_y}{p_z} & |p_y|^2 & \inner{p_y}{p_z} \\
    \inner{p_z}{p_x} & \inner{p_z}{p_y} & |p_z|^2 \\
  \end{bmatrix}
  =
  \frac{1}{4}
  \begin{bmatrix}
    5 & 0 & 0 \\
    0 & 2 & -1 \\
    0 & -1 & 2 \\
  \end{bmatrix}, 
\end{equation}
Take $\v{p}_x$, $\v{p}_y$, and $\v{p}_z$
\begin{equation}
  \label{eq:hcp:lattice}
  \v{p}_x
  =
  \left(\sqrt{5}, 0, 0\right), 
  \quad
  \v{p}_y
  =
  \left(1, \frac{1}{\sqrt{2}}, \sqrt{\frac{3}{2}}\right), 
  \quad
  \v{p}_z
  =
  \left(0, \frac{1}{\sqrt{2}}, -\sqrt{\frac{3}{2}}\right), 
\end{equation}
which satisfy Equation (\ref{eq:hcp:gram}), 
and substituting Equation (\ref{eq:hcp:lattice}) into Equation (\ref{eq:hcp:vector}), 
we obtain vectors of edges as

\begin{displaymath}
  \begin{alignedat}{3}
    \v{e}_1 &= \left(\frac{\sqrt{5}}{2}, 0, 0\right), 
    &\quad
    \v{e}_2 &= \left(-\frac{\sqrt{5}}{2}, 0, 0\right), 
    \\
    \v{e}_3 &= \left(-\frac{1}{2 \sqrt{5}}, -\frac{1}{\sqrt{2}}, -\frac{1}{\sqrt{6}}\right), 
    &\quad
    \v{e}_4 &= \left(-\frac{1}{2 \sqrt{5}}, \frac{1}{\sqrt{2}}, -\frac{1}{\sqrt{6}}\right), 
    \\
    \v{e}_5 &= \left(-\frac{1}{2 \sqrt{5}}, 0, \sqrt{\frac{2}{3}}\right), 
    &\quad
    \v{e}_6 &= \left(-\frac{3}{2 \sqrt{5}}, 0, 0\right), 
    \\
    \v{e}_7 &= \left(-\frac{1}{2 \sqrt{5}}, \frac{1}{\sqrt{2}}, \frac{1}{\sqrt{6}}\right), 
    &\quad
    \v{e}_8 &= \left(-\frac{1}{2 \sqrt{5}}, 0, -\sqrt{\frac{2}{3}}\right), 
    \\
    \v{e}_9 &= \left( -\frac{1}{2 \sqrt{5}}, -\frac{1}{\sqrt{2}}, \frac{1}{\sqrt{6}}\right), 
    &\quad
    \v{e}_{10} &= \left( \frac{1}{2 \sqrt{5}}, -\frac{1}{\sqrt{2}}, \frac{1}{\sqrt{6}}\right), 
    \\
    \v{e}_{11} &= \left(\frac{1}{2 \sqrt{5}}, \frac{1}{\sqrt{2}}, \frac{1}{\sqrt{6}}\right), 
    &\quad
    \v{e}_{12} &= \left(\frac{1}{2 \sqrt{5}}, 0, -\sqrt{\frac{2}{3}}\right), 
    \\
    \v{e}_{13} &= \left(\frac{3}{2 \sqrt{5}}, 0, 0\right), 
    &\quad
    \v{e}_{14} &= \left(\frac{1}{2 \sqrt{5}}, \frac{1}{\sqrt{2}}, -\frac{1}{\sqrt{6}}\right), 
    \\
    \v{e}_{15} &= \left(\frac{1}{2 \sqrt{5}}, -\frac{1}{\sqrt{2}}, -\frac{1}{\sqrt{6}}\right), 
    &\quad
    \v{e}_{16} &= \left(\frac{1}{2 \sqrt{5}}, 0, \sqrt{\frac{2}{3}}\right)
  \end{alignedat}
\end{displaymath}
and positions of vertices as
\begin{displaymath}
  \begin{alignedat}{4}
    \v{v}_0 &= (0, 0, 0), 
    \\
    \v{v}_1 &= \v{v}_0 + \v{e}_1 
    = \left(\frac{\sqrt{5}}{2}, 0, 0\right)
    &\quad
    \v{v}_2 &= \v{v}_0 - \v{e}_7
    =  \left(\frac{1}{2 \sqrt{5}}, -\frac{1}{\sqrt{2}}, -\frac{1}{\sqrt{6}}\right), 
    \\
    \v{v}_3 &= \v{v}_1 + \v{e}_3
    = \left(-\frac{1}{2 \sqrt{5}}, -\frac{1}{\sqrt{2}}, -\frac{1}{\sqrt{6}}\right)
    &\quad
    \v{v}_4 &= \v{v}_0 - \v{e}_{14}
    = \left(-\frac{1}{2 \sqrt{5}}, -\frac{1}{\sqrt{2}}, \frac{1}{\sqrt{6}}\right)
    \\
    \v{v}_5 &= \v{v}_1 + \v{e}_{10}
    = \left(\frac{1}{2 \sqrt{5}}, -\frac{1}{\sqrt{2}}, \frac{1}{\sqrt{6}}\right)
    &\quad
    \v{v}_{0a} &= \v{v}_4 + \v{e}_{15}
    = \left(0, -\sqrt{2}, 0\right), 
    \\
    \v{v}_{0b} &= \v{v}_4 + \v{e}_{16}
    = \left(0, -\frac{1}{\sqrt{2}}, \sqrt{\frac{3}{2}}\right)
    &\quad
    \v{v}_{0c} &= \v{v}_2 + \v{e}_{8}
    = \left(0, -\frac{1}{\sqrt{2}}, -\sqrt{\frac{3}{2}}\right)
    \\
    \v{v}_{0d} &= \v{v}_2 + \v{e}_{9}
    = \left(0, -\sqrt{2}, 0\right)
    &\quad
    \v{v}_{1a} &= \v{v}_0 + \v{e}_{2}
    = \left(-\frac{\sqrt{5}}{2}, 0, 0\right)
    \\
    \v{v}_{2a} &= \v{v}_3 + \v{e}_{6}
    = \left(-\frac{2}{\sqrt{5}}, -\frac{1}{\sqrt{2}}, -\frac{1}{\sqrt{6}}\right)
    &\quad
    \v{v}_{3a} &= \v{v}_1 + \v{e}_{4}
    = \left(\frac{2}{\sqrt{5}}, \frac{1}{\sqrt{2}}, -\frac{1}{\sqrt{6}}\right)
    \\
    \v{v}_{3b} &= \v{v}_1 + \v{e}_{5}
    = \left(\frac{2}{\sqrt{5}}, 0, \sqrt{\frac{2}{3}}\right)
    &\quad
    \v{v}_{4a} &= \v{v}_5 + \v{e}_{13}
    = \left(\frac{2}{\sqrt{5}}, -\frac{1}{\sqrt{2}}, \frac{1}{\sqrt{6}}\right)
    \\
    \v{v}_{5a} &= \v{v}_1 + \v{e}_{12}
    = \left(\frac{3}{\sqrt{5}}, 0, -\sqrt{\frac{2}{3}}\right)
    &\quad
    \v{v}_{5b} &= \v{v}_1 + \v{e}_{11}
    = \left(\frac{3}{\sqrt{5}}, \frac{1}{\sqrt{2}}, \frac{1}{\sqrt{6}}\right)
  \end{alignedat}
\end{displaymath}

The dual crystal structure and crystal structure obtained  from this graph are shown in Figure \ref{fig:hcp:structure}c and d. 
The obtained crystal structure is consistent with the HCP structure. 
Therefore, the HCP structure can be reproduced from the graph using our method.
\subsection{Crystal Structure Generation for BCC Structures}
\begin{figure}[hbtp]
  \centering
  \includegraphics[width=1\textwidth]{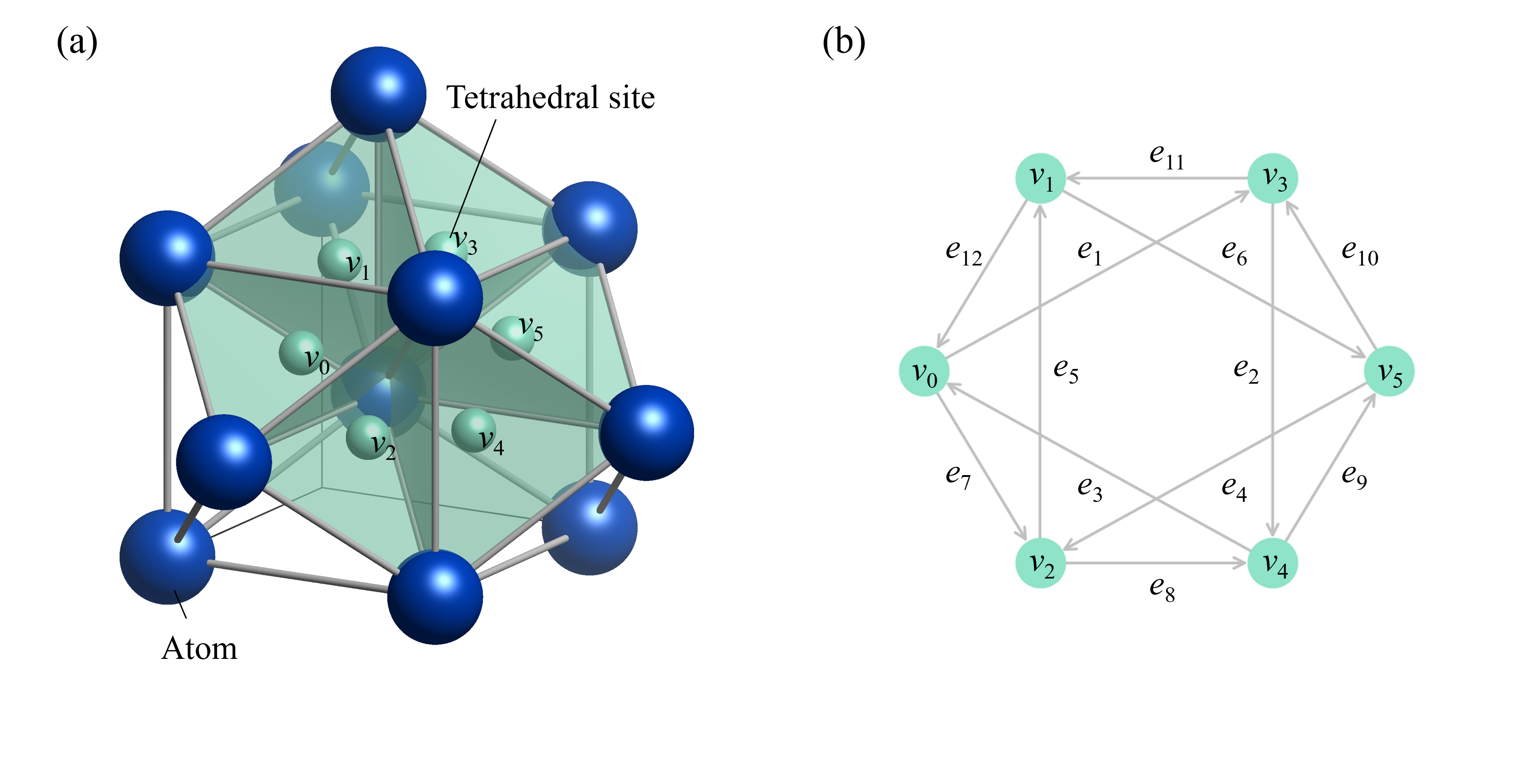}
  \includegraphics[width=1\textwidth]{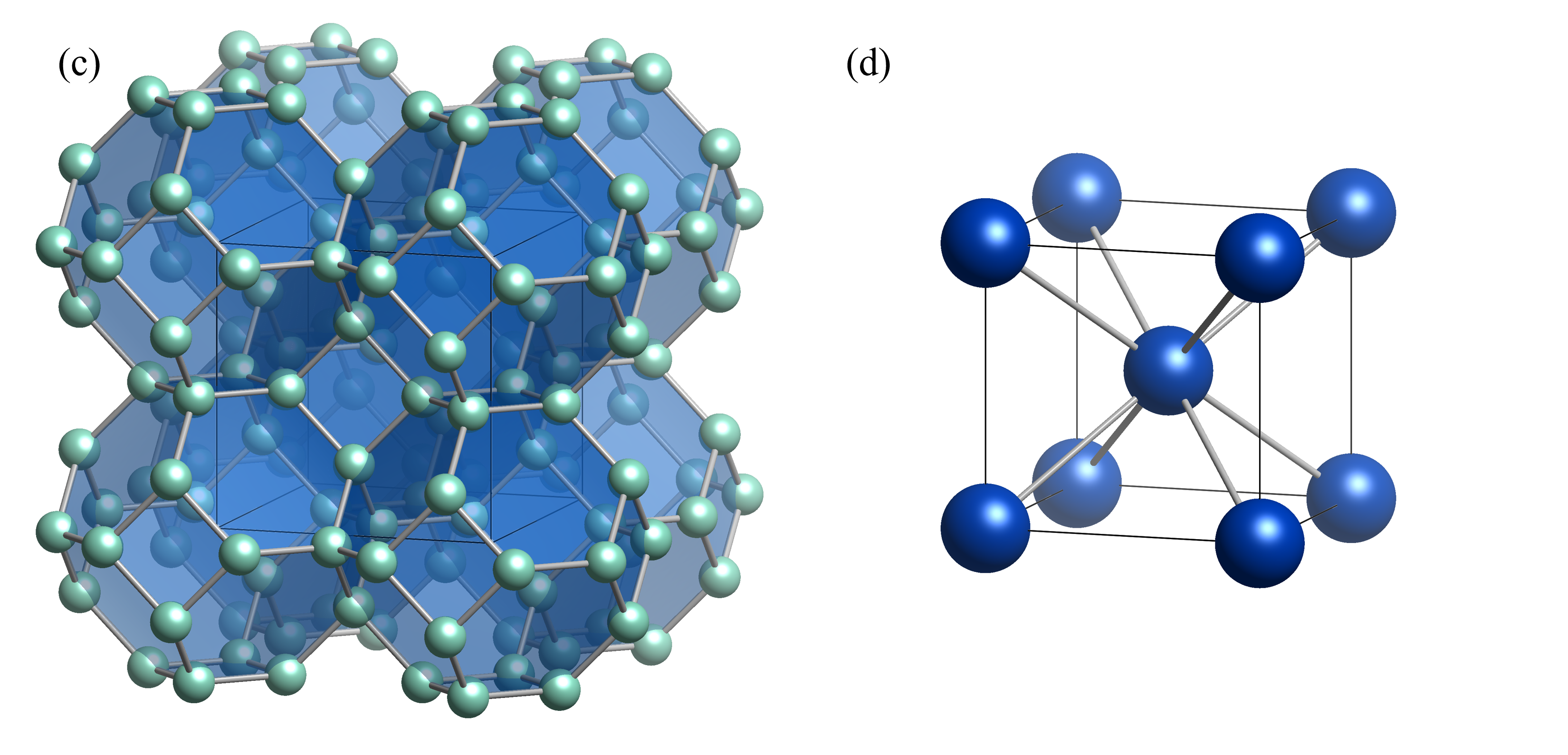}
  \caption{(a) Primitive cell of the BCC structure consisting of six tetrahedra. (b) Dual periodic graph of the BCC structure, where ${v_i}$ and ${e_i}$ are the vertex and edge numbers, respectively. (c) Dual crystal structure and (d) crystal structure generated from the dual periodic graph using our method.}
  \label{fig:bcc:structure}
\end{figure}

Next, we demonstrate crystal structure generation from dual periodic graphs of BCC structures using our method. 
The BCC structure and dual periodic graph are shown in Figure \ref{fig:bcc:structure}a and b, respectively. 
The primitive BCC structure has six interstitial tetrahedral sites. Unlike the FCC and HCP structures, the BCC structure is tessellated from only tetrahedra.
The dual periodic graph has four edges connecting to the red vertex, corresponding to the face number of tetrahedra.

The specific process of generating the BCC structure from its dual periodic graph using standard realization theory is described below. 
The dual periodic graph $X_0$ of BCC structure had $b = 7$.
A $\Z$-basis of $H_1(X_0, \Z)$ was taken as
\begin{displaymath}
  \begin{alignedat}{3}    
    \alpha_1 &= e_1 + e_{11} + e_{12}, 
    &\quad
    \alpha_2 &= e_2 + e_9 + e_{10}, 
    &\quad
    \alpha_3 &= e_3 + e_7 + e_8, 
    \\
    \alpha_4 &= e_5 - e_8 + e_2 - e_{11}, 
    &\quad
    \alpha_5 &= e_6 - e_9 + e_3 - e_{12}, 
    &\quad
    \alpha_6 &= e_7 - e_4- e_1 + e_{10}, 
    \\
    \alpha_7 &= e_7 + e_8 + e_9 + e_{10} + e_{11} + e_{12}, 
  \end{alignedat}
\end{displaymath}
The closed path corresponding to each basis is shown in Figure S3.
Then, projecting onto the subspace spanning $\alpha_1$, $\alpha_2$, and $\alpha_3$, 
we obtain a standard realization in $3$D space as 
\begin{equation}
  \label{eq:bcc:vector}
  \begin{alignedat}{3}
    e_1 &= \frac{1}{2} p_x + \frac{1}{4} p_y + \frac{1}{4} p_z, 
    &\quad
    e_2 &= \frac{1}{4} p_x + \frac{1}{2} p_y + \frac{1}{4} p_z, 
    \\
    e_3 &= \frac{1}{4} p_x + \frac{1}{4} p_y + \frac{1}{2} p_z,
    &\quad
    e_4 &= -\frac{1}{2} p_x -\frac{1}{4} p_y - \frac{1}{4} p_z,
    \\
    e_5 &= -\frac{1}{4} p_x -\frac{1}{2} p_y - \frac{1}{4} p_z, 
    &\quad
    e_6 &= -\frac{1}{4} p_x -\frac{1}{4} p_y - \frac{1}{2} p_z, 
    \\
    e_7 &= -\frac{1}{4} p_y + \frac{1}{4} p_z, 
    &\quad
    e_8 &= -\frac{1}{4} p_x + \frac{1}{4} p_z, 
    \\
    e_9 &= \frac{1}{4} p_x + \frac{1}{4} p_y, 
    &\quad
    e_{10} &= \frac{1}{4} p_y -\frac{1}{4} p_z,
    \\
    e_{11} &= \frac{1}{4} p_x -\frac{1}{4} p_z, 
    &\quad
    e_{12} &= \frac{1}{4} p_x -\frac{1}{4} p_y
  \end{alignedat}
\end{equation}
The period vectors (a basis of a Bravais lattice) $\{\v{p}_x, \v{p}_y, \v{p}_z\}$ satisfy \begin{equation}
  \label{eq:bcc:gram}
  G
  =
  \begin{bmatrix}
    |p_x|^2 & \inner{p_x}{p_y} & \inner{p_x}{p_z} \\
    \inner{p_y}{p_z} & |p_y|^2 & \inner{p_y}{p_z} \\
    \inner{p_z}{p_x} & \inner{p_z}{p_y} & |p_z|^2 \\
  \end{bmatrix}
  =
  \begin{bmatrix}
    3 & -1 & -1 \\
    -1 & 3 & -1 \\
    -1 & -1 & 3 \\
  \end{bmatrix}.
\end{equation}
Take $\v{p}_x$, $\v{p}_y$, and $\v{p}_z$
\begin{equation}
  \label{eq:bcc:lattice}
  \v{p}_x
  =
  \left(-1, 1, 1\right), 
  \quad
  \v{p}_y
  =
  \left(1, -1, 1\right), 
  \quad
  \v{p}_z
  =
  \left(1, 1, -1\right), 
\end{equation}
which satisfy Equation (\ref{eq:bcc:gram}), 
and substituting Equation (\ref{eq:bcc:lattice}) into Equation (\ref{eq:bcc:vector}), 
we obtain vectors of edges as
\begin{displaymath}
  \begin{alignedat}{4}
    \v{e}_1 &= \left(0, \frac{1}{2}, \frac{1}{2}\right), 
    &\quad
    \v{e}_2 &= \left(\frac{1}{2}, 0, \frac{1}{2}\right), 
    \\
    \v{e}_3 &= \left(\frac{1}{2}, \frac{1}{2}, 0\right), 
    &\quad
    \v{e}_4 &= \left(0, -\frac{1}{2}, -\frac{1}{2}\right), 
    \\
    \v{e}_5 &= \left(-\frac{1}{2}, 0, -\frac{1}{2}\right), 
    &\quad
    \v{e}_6 &= \left(-\frac{1}{2}, -\frac{1}{2},  0\right), 
    \\
    \v{e}_7 &= \left(0, \frac{1}{2}, -\frac{1}{2}\right), 
    &\quad
    \v{e}_8 &= \left(\frac{1}{2}, 0, -\frac{1}{2}\right), 
    \\
    \v{e}_9 &= \left(\frac{1}{2}, -\frac{1}{2}, 0\right), 
    &\quad
    \v{e}_{10} &= \left(0, -\frac{1}{2}, \frac{1}{2}\right), 
    \\
    \v{e}_{11} &= \left(-\frac{1}{2}, 0, \frac{1}{2}\right), 
    &\quad
    \v{e}_{12} &= \left(-\frac{1}{2}, \frac{1}{2}, 0\right)
  \end{alignedat}
\end{displaymath}
and positions of vertices as
\begin{displaymath}
  \begin{alignedat}{4}
    \v{v}_0 &= (0, 0, 0), 
    \\
    \v{v}_1 &= \v{v}_0 + \v{e}_1 
    =
    \left(0, \frac{1}{2}, \frac{1}{2}\right)
    &\quad
    \v{v}_2 &= \v{v}_0 - \v{e}_7
    =
    \left(0, -\frac{1}{2}, \frac{1}{2}\right)
    \\
    \v{v}_3 &= \v{v}_1 + \v{e}_3
    = \left(\frac{1}{2}, \frac{1}{2}, 0\right)
    &\quad
    \v{v}_4 &= \v{v}_0 - \v{e}_{12}
    = \left(\frac{1}{2}, -\frac{1}{2}, 0\right)
    \\
    \v{v}_5 &= \v{v}_1 + \v{e}_{10}
    = \left(0, -\frac{1}{2}, \frac{1}{2}\right)
    &\quad
    \v{v}_{1a} &= \v{v}_2 + \v{e}_5
    = \left(-\frac{1}{2}, -\frac{1}{2}, 0\right)
    \\
    \v{v}_{1b} &= \v{v}_3 + \v{e}_{11}
    = \left(0, \frac{1}{2}, \frac{1}{2}\right)
    &\quad
    \v{v}_{2a} &= \v{v}_5 + \v{e}_4
    = \left(0, -1, 0\right)
    \\
    \v{v}_{4a} &= \v{v}_3 + \v{e}_2
    = \left(1, \frac{1}{2}, \frac{1}{2}\right)
    &\quad
    \v{v}_{4b} &= \v{v}_2 + \v{e}_8
    = \left(\frac{1}{2}, -\frac{1}{2}, 0\right)
    \\
    \v{v}_{5a} &= \v{v}_1 + \v{e}_6
    = \left(-\frac{1}{2}, 0, \frac{1}{2}\right)
    &\quad
    \v{v}_{5b} &= \v{v}_3 + \v{e}_9
    = \left(1, 0, 0\right)
  \end{alignedat}
\end{displaymath}

The dual crystal structure and crystal structure obtained from this dual periodic graph are shown in Figure \ref{fig:bcc:structure}c and d, respectively. 
The obtained crystal structure is consistent with the BCC structure. 
Therefore, the BCC structure can be reproduced from the dual periodic graph using our method.

\subsection{Challenges of Structure Generation by Standard Realizations}
In this work, we generated FCC, HCP, and BCC structures from each corresponding dual periodic graph using our calculation method. 
Our method can be applied to generate undiscovered crystal structures from target polyhedra if a dual periodic graph is obtained from the type and number of the polyhedra.
For example, by generating crystal structures using only combinatorial information of tetrahedra as input, it is possible to generate new tetrahedrally packed structures other than BCC structures.
However, for more complex structures, 
it is not easy to select an appropriate basis of closed paths from a huge number of combinations of closed paths.
To generate crystal structures from graphs using standard realization theory, we should select an appropriate $3$D vector space from the $b$-dimensional vector space.
In addition, it is necessary to select appropriate closed paths that define the $(b-3)$-dimensional subspace for the basis of the closed paths.

For example, in a dual periodic graph with a BCC structure, there are eight closed paths consisting of three nodes (Figure S4), 15 closed paths consisting of four nodes (Figure S5), 24 closed paths consisting of five nodes (Figure S6), and 16 closed paths consisting of six nodes (Figure S7), giving a total of 63 closed paths. 
To generate the dual BCC structure, seven closed paths corresponding to $b$ must be selected.
We consider three closed paths consisting of three nodes, three consisting of four nodes, and one consisting of six nodes as the seven closed paths.
The number of such closed path combinations is $\binom{8}{3}$ $\times$ $\binom{15}{3}$ $\times$ $\binom{16}{1}$ (=407,680).
Among them, we have identified 128 combinations of closed paths that can generate the structure of the most symmetric $Im\overline{3}m$ space group. 
It is noted that the combination of closed paths in Figure S3 is a representative example of these 128 combinations.
Other combinations of closed paths either generated crystal structures that are less symmetric than the $Im\overline{3}m$ space group or failed to generate crystal structures. 
We found the closed path combination shown in Figure S3 by trial and error, but the rule for selecting the appropriate closed path combination is still unclear. 
Even for structures with a small number of atoms in the unit cell and high symmetry, such as BCC structures, it is difficult to generate the correct structure because of the large number of closed path combinations. This problem needs to be solved to be able to apply this method to more complex crystal structures. In the future, we plan to apply this method to more complex crystal structures and develop a method to efficiently select the appropriate closed path combination.

When using this method to predict an unknown structure, it is necessary to define a dual periodic graph from the given polyhedra. 
Given a set of polyhedra, the number of vertices and the number of edges joining between vertices in the dual periodic graph are determined, and the dual periodic graph can be defined by determining which edges are connected to each other. 
For example, a dual periodic graph can be obtained by determining the connectable faces between polyhedra based on the given shape of the polyhedra. 
By applying our method to all combinations of such obtained dual periodic graphs, it is possible, in principle, to generate crystal structures from any polyhedron.

Our structure generation method, when integrated with other techniques, should allow the prediction of practical structures. 
For instance, combining our approach with the atomic-site configuration optimization method should enable the prediction of complex crystal structures composed of various elemental species. 
Recent reports indicate that quantum annealing can be used to solve atomic-site optimization problems at low cost \cite{Ichikawa2023}, facilitating the prediction of existing multicomponent compounds \cite{Choubisa2023}.

The integration of our method with phonon analysis can be used to predict low-symmetry structures. 
Our method generates the most symmetrical structure for a given graph, so it cannot produce structures with low symmetry. 
Conducting a phonon calculation of the derived structure, using either first-principles calculations or machine learning potentials, can generate a low-symmetry structure along the direction of structural instability \cite{Togo2013}. 
These optimization methods require the input of the target crystal structure, so their combination with our crystal structure generation method can facilitate exploration of crystalline materials.

Finally, we compare our method based on discrete geometry with crystal structure prediction methods by generative models such as GAN, VAE, and diffusion models. 
While a large amount of training data is required to accurately predict crystal structures with generative models \cite{Zeni2023}, our method, which does not use machine learning techniques, needs no training data.
In the generative model approach, the lattice vectors and atomic coordinates of the crystal structure are predicted by independent models or processes \cite{Noh2019}. 
In contrast, our method can directly transform graphs into corresponding lattice vectors and atomic coordinates in a unified process. 
A notable limitation of generative models is that they tend to predict structures with low symmetry \cite{Zeni2023}.
Our approach, however, consistently generates structures with high symmetry. 
The inherent capability of our method to facilitate the bidirectional transformation between graphs and crystal structures presents a promising avenue for the discovery of new crystal structures that would be difficult with conventional methods, for example when it is employed in conjunction with a generative model designed to generate graphs.

\section{Conclusion}
In this study, we proposed a method for crystal structure generation based on the discrete geometry of target polyhedra.
Most conventional methods of crystal structure prediction use random structure generation, which produces many unimportant structures and contributes to inefficiency in material exploration.
In the developed method, the shape and connectivity of space-filling polyhedra are represented as a dual periodic graph and crystal structures are generated from this graph based on standard realization theory.
We demonstrated that this method can generate FCC, HCP, and BCC structures from dual periodic graphs.
This work is a first step toward generating undiscovered crystal structures based on the target polyhedra.
Thus, this work paves the way for structure-driven materials development and has the potential to identify highly functional materials that have been difficult to discover with conventional composition-driven development.



\begin{suppinfo}

Basis for closed paths of FCC, HCP, and BCC structures used in this study, all closed paths of BCC structures, and example of python code to generate a dual structure.

\end{suppinfo}

\bibliography{ref}

\end{document}


\begin{figure}[hbtp]
  \centering
  \includegraphics[width=1\textwidth]{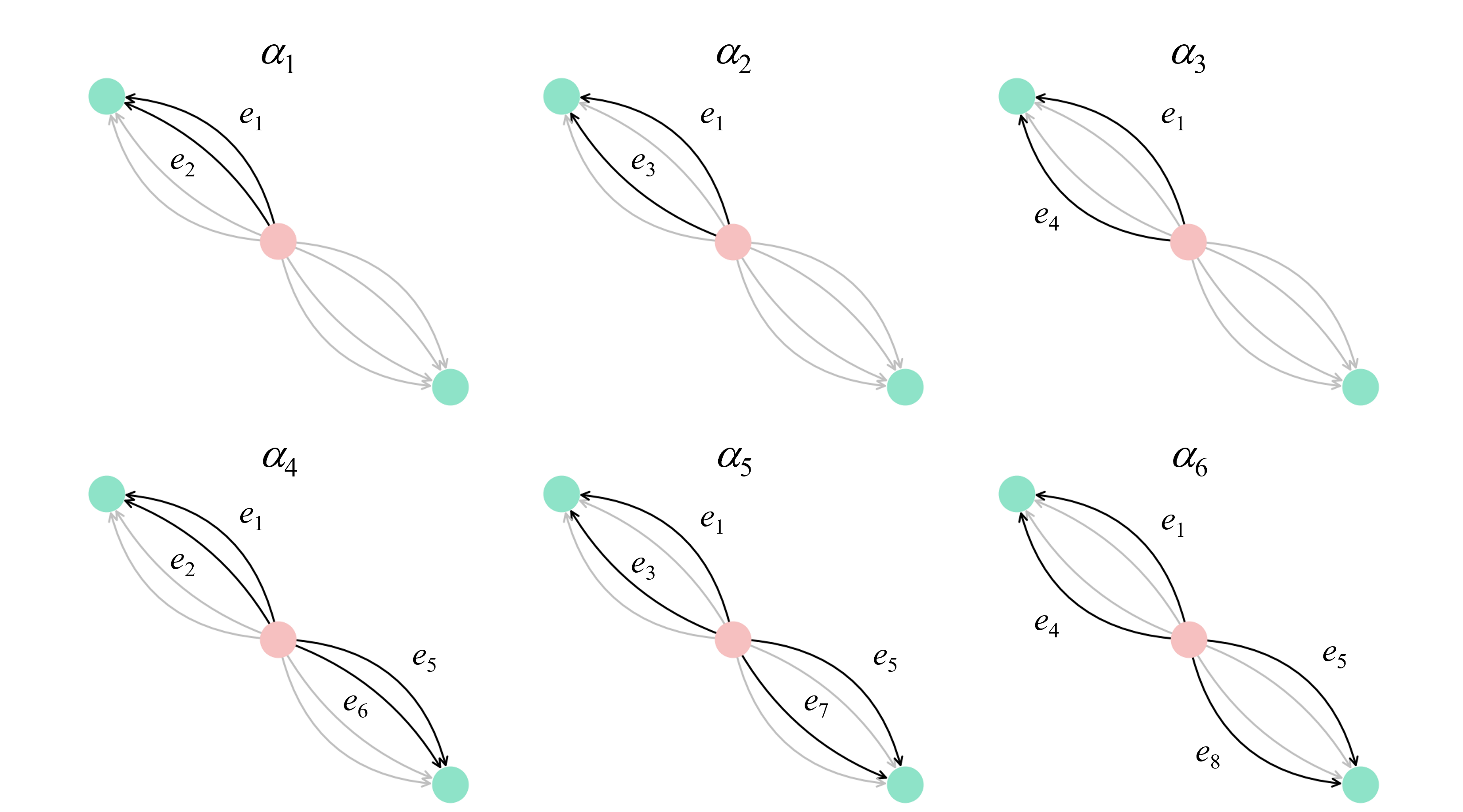}
  \caption{Basis of closed paths of FCC structure}
  \label{fig:fcc:basis}
\end{figure}

\begin{figure}[hbtp]
  \centering
  \includegraphics[width=1\textwidth]{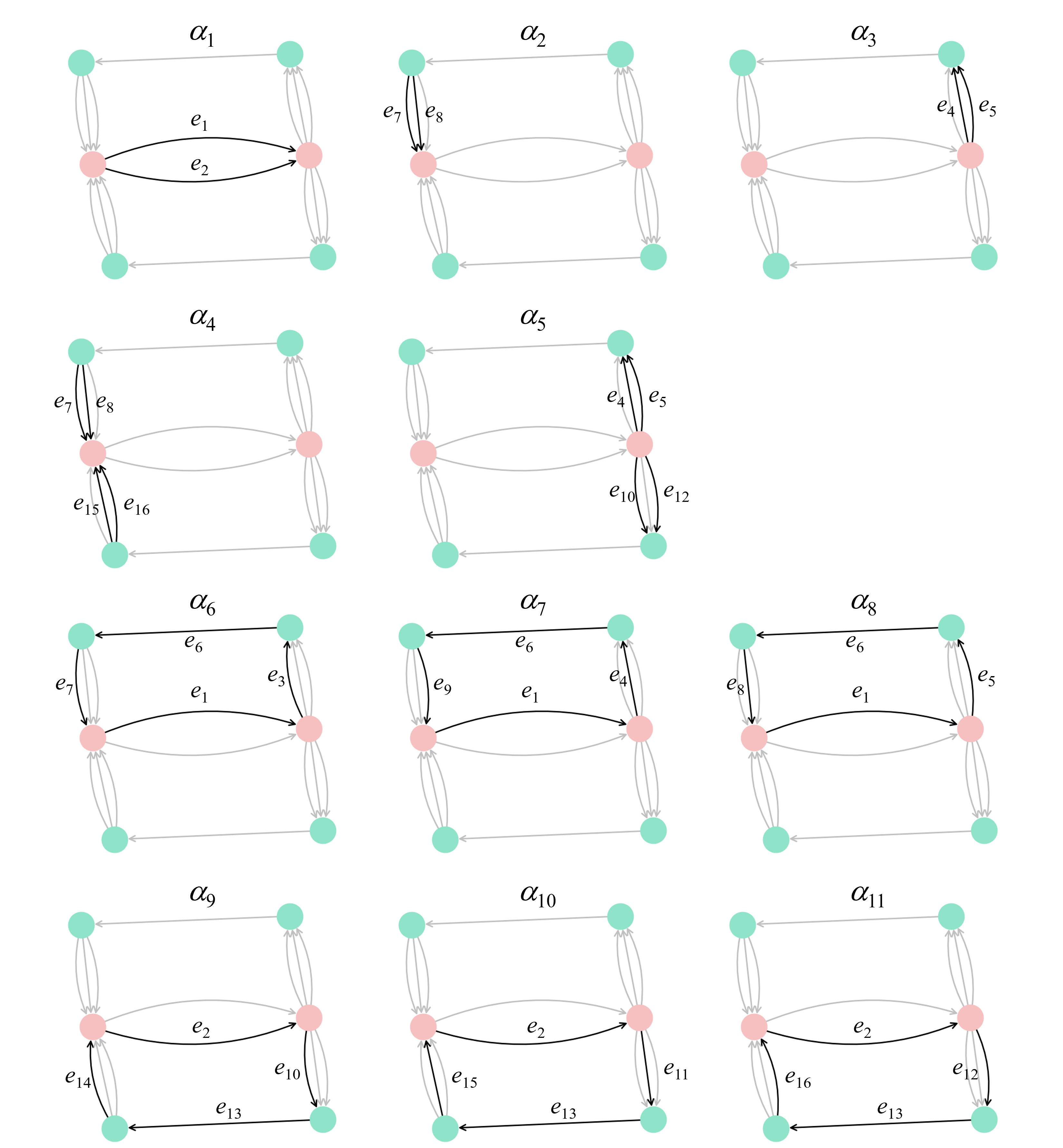}
  \caption{Basis of closed paths of HCP structure}
  \label{fig:hcp:basis}
\end{figure}

\begin{figure}[hbtp]
  \centering
  \includegraphics[width=1\textwidth]{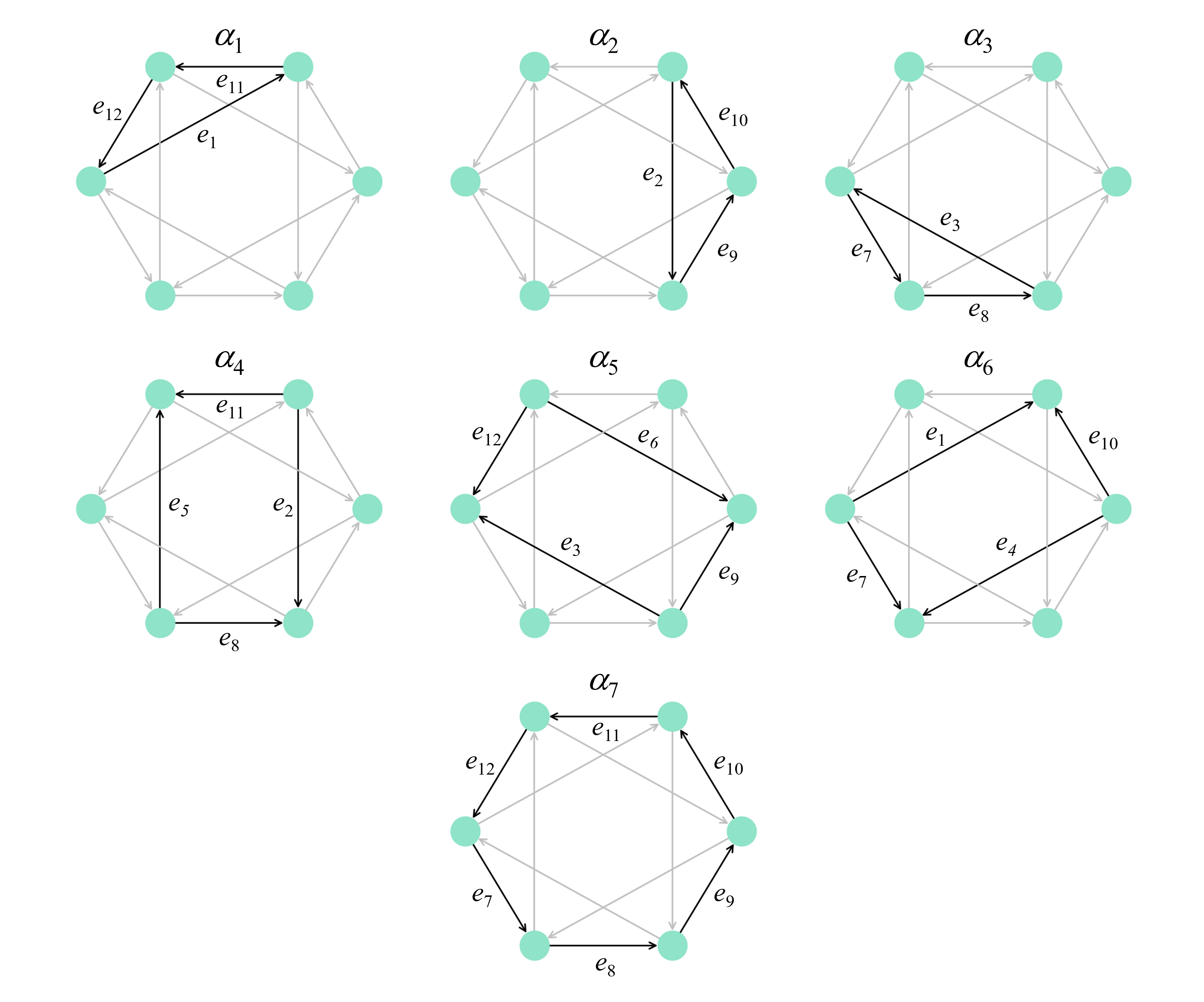}
  \caption{Basis of closed paths of BCC structure}
  \label{fig:bcc:basis}
\end{figure}

\begin{figure}[hbtp]
  \centering
  \includegraphics[width=1\textwidth]{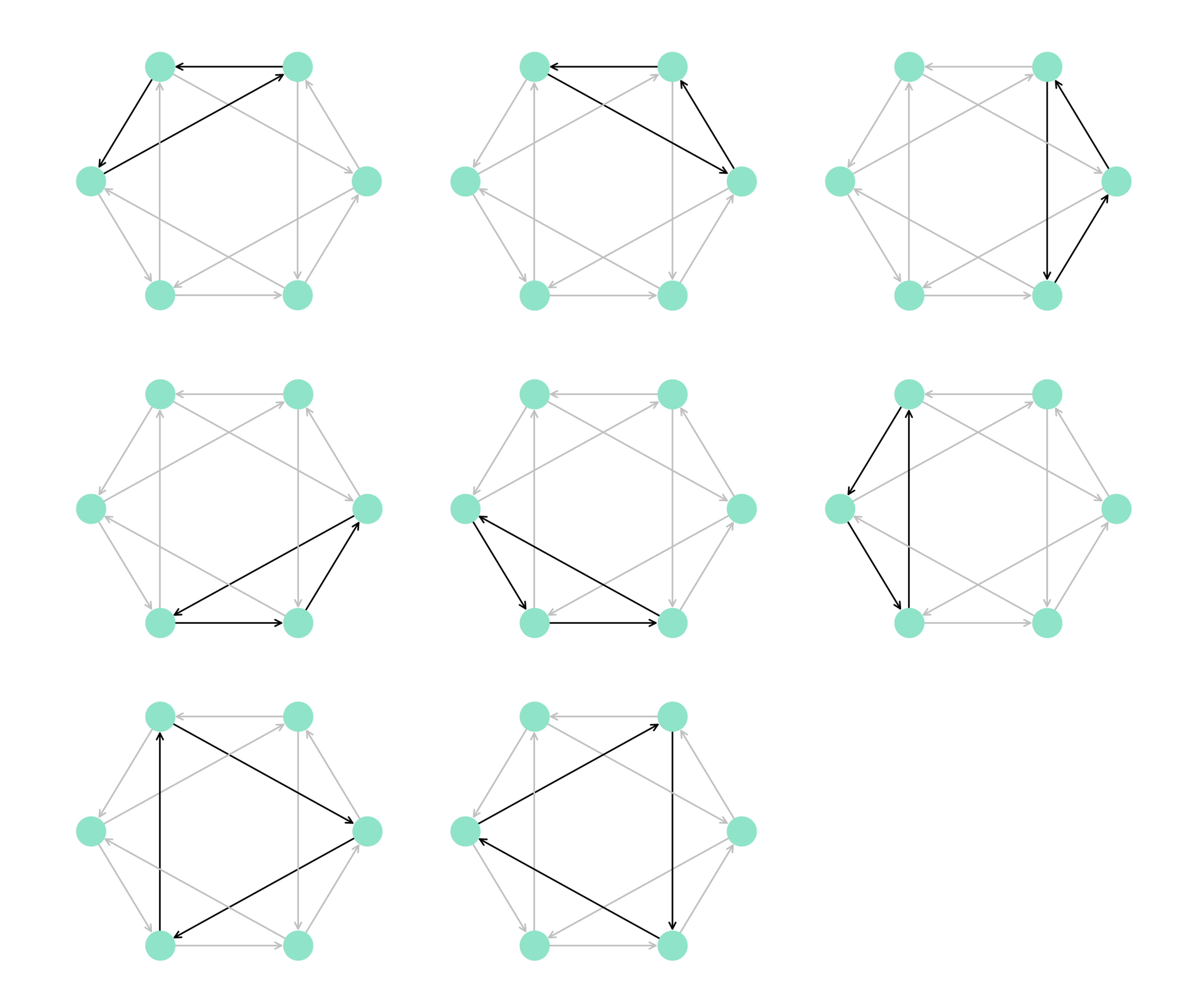}
  \caption{Closed paths consisting of three nodes for BCC structure}
  \label{fig:bcc:3node:path}
\end{figure}

\begin{figure}[hbtp]
  \centering
  \includegraphics[width=1\textwidth]{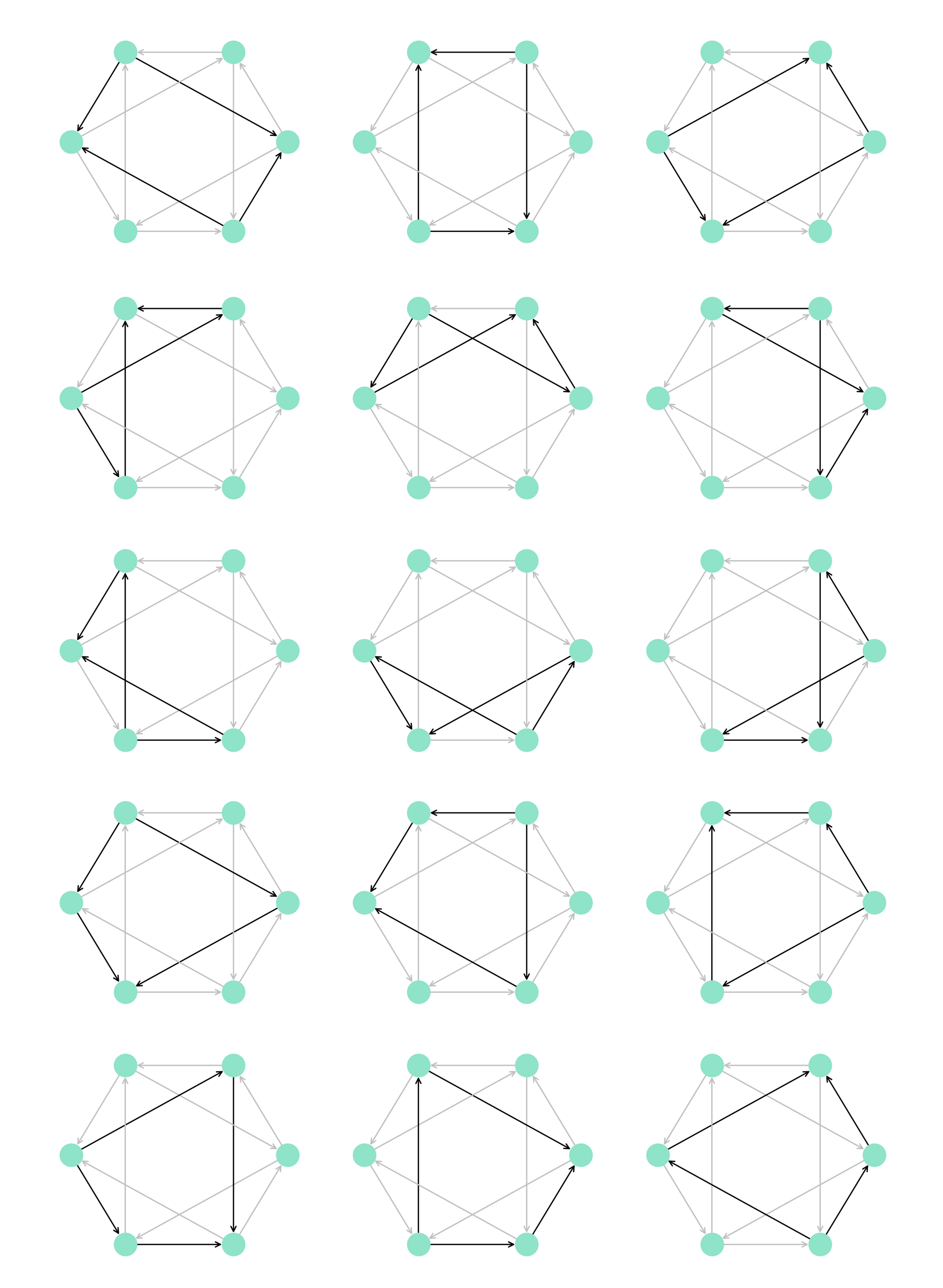}
  \caption{Closed paths consisting of four nodes for BCC structure}
  \label{fig:bcc:4node:path}
\end{figure}

\begin{figure}[hbtp]
  \centering
  \includegraphics[width=1\textwidth]{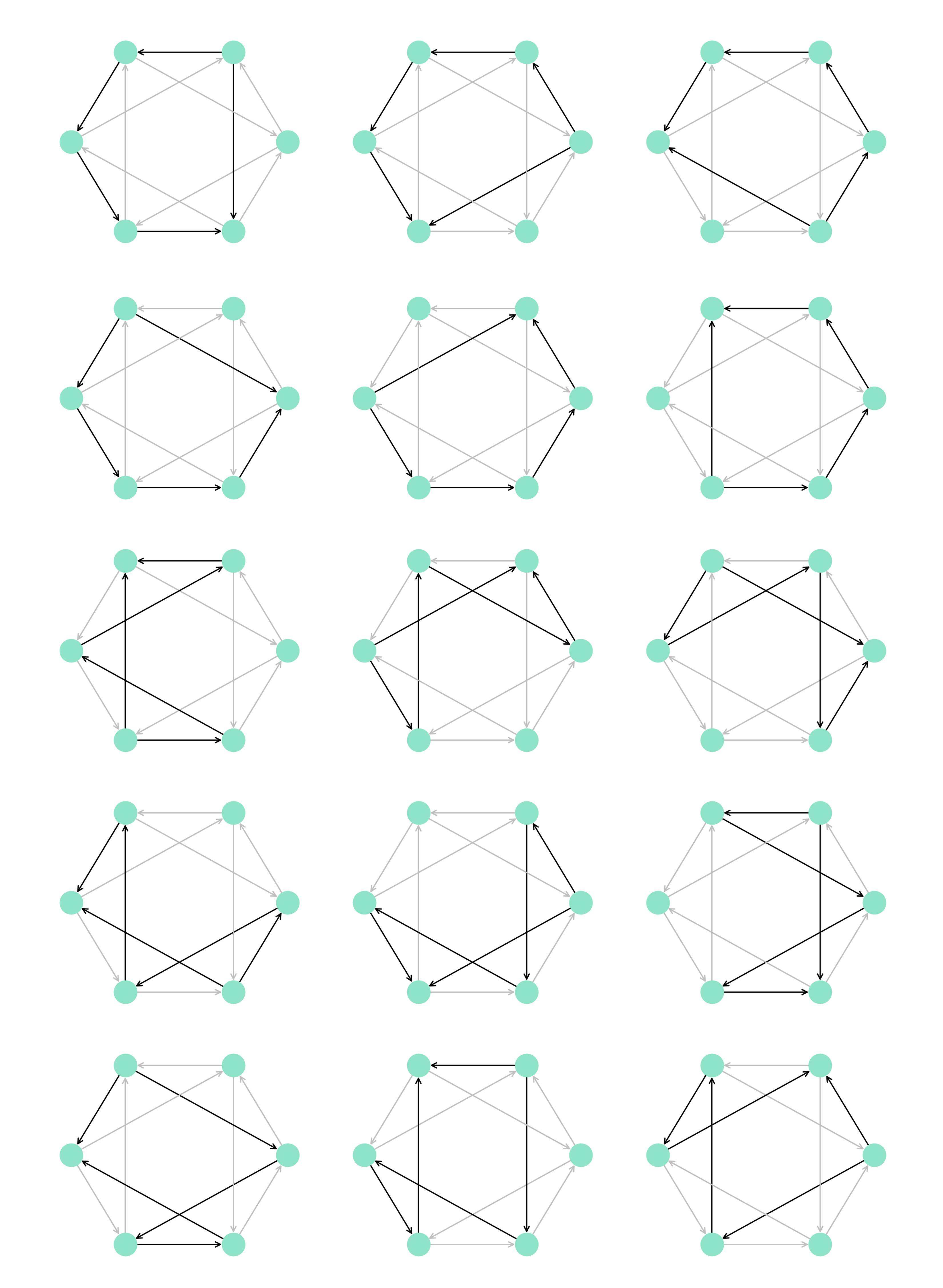}
  \label{fig:bcc:5node:path:a}
\end{figure}
\begin{figure}[hbtp]
  \centering
  \includegraphics[width=1\textwidth]{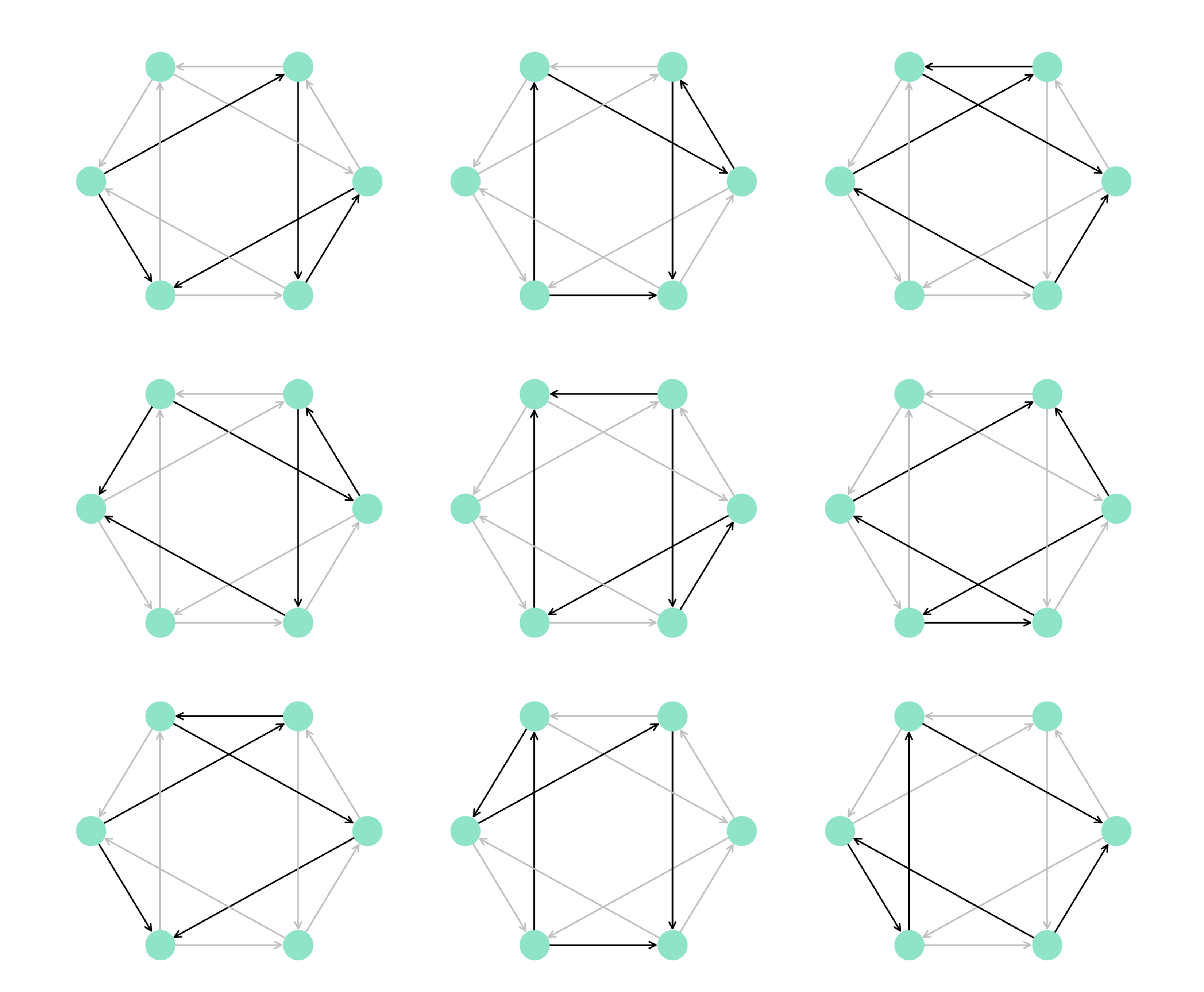}
  \caption{Closed paths consisting of five nodes for BCC structure}
  \label{fig:bcc:5node:path:b}
\end{figure}

\begin{figure}[hbtp]
  \centering
  \includegraphics[width=1\textwidth]{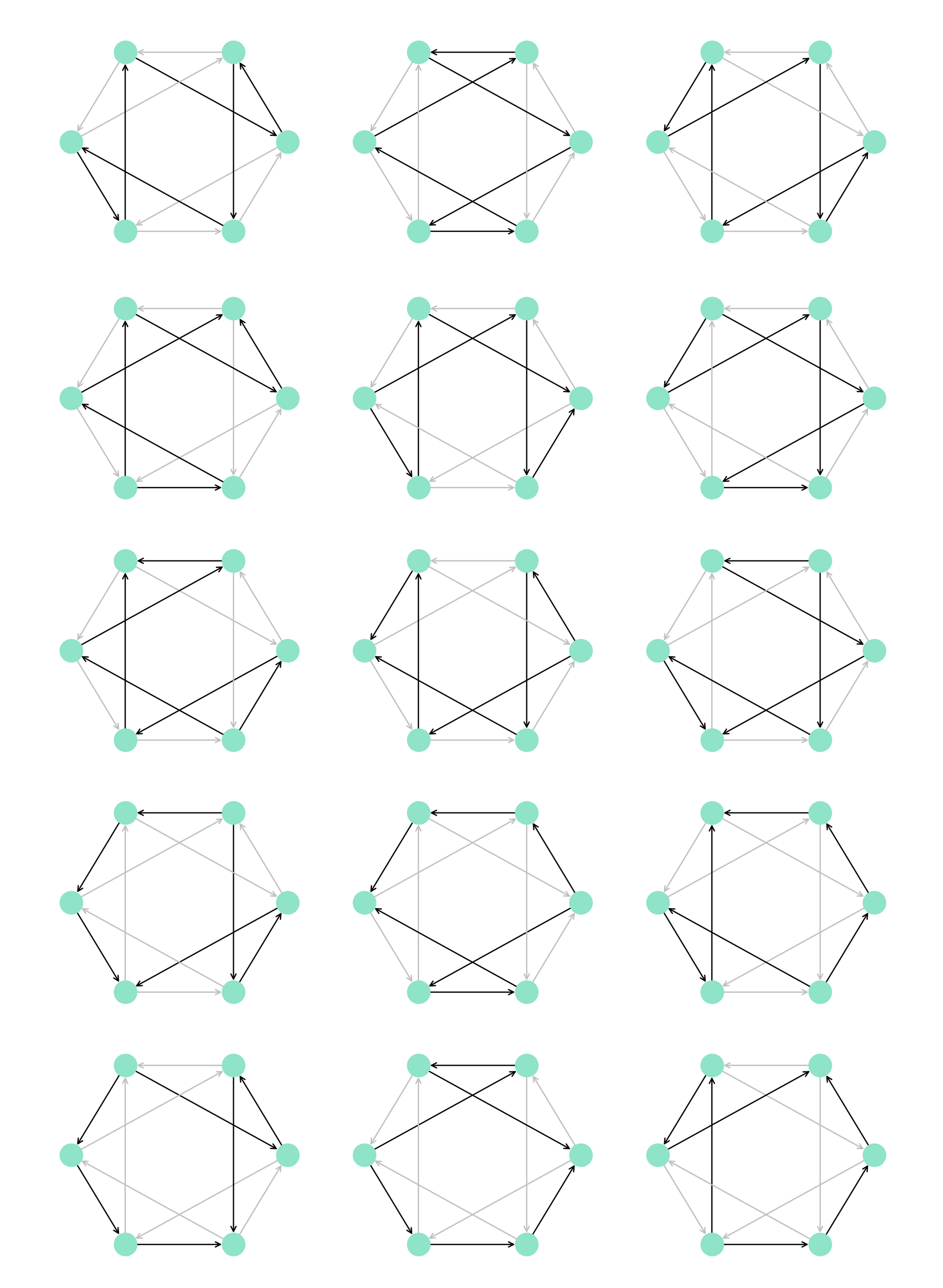}
  \label{fig:bcc:6node:path:a}
\end{figure}

\begin{figure}[hbtp]
  \centering
  \includegraphics[width=1\textwidth]{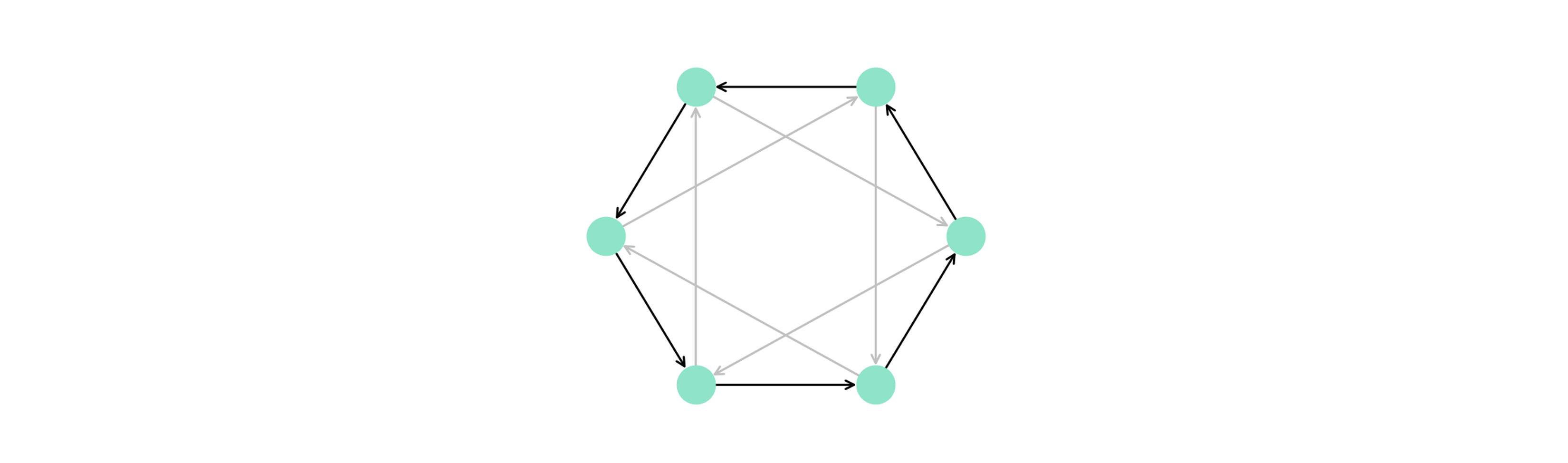}
  \caption{Closed paths consisting of six nodes for BCC structure}
  \label{fig:bcc:6node:path:b}
\end{figure}

\begin{lstlisting}[language=Python, caption=Example of Python code to generate a dual structure]
import networkx as nx
import numpy as np
from pymatgen.core import Structure, Element, Lattice

def set_graph(struct_type):
    graph = nx.MultiDiGraph()
    if struct_type == "fcc":
        graph.add_node("v0")
        graph.add_node("v1")
        graph.add_node("v2")
        graph.add_edge("v0", "v1", 0) #e1
        graph.add_edge("v0", "v1", 1) #e2
        graph.add_edge("v0", "v1", 2) #e3
        graph.add_edge("v0", "v1", 3) #e4
        graph.add_edge("v0", "v2", 4) #e5
        graph.add_edge("v0", "v2", 5) #e6
        graph.add_edge("v0", "v2", 6) #e7
        graph.add_edge("v0", "v2", 7) #e8

        ne = len(graph.edges)
        e = np.identity(ne)
        alpha = np.array([
            e[0] - e[1],                #alpha1
            e[0] - e[2],                #alpha2
            e[0] - e[3],                #alpha3
            e[0] - e[1] + e[4] - e[5],  #alpha4
            e[0] - e[2] + e[4] - e[6],  #alpha5
            e[0] - e[3] + e[4] - e[7]   #alpha6
        ])
    elif struct_type == "hcp":
        graph.add_node("v0")
        graph.add_node("v1")
        graph.add_node("v2")
        graph.add_node("v3")
        graph.add_node("v4")
        graph.add_node("v5")
        graph.add_edge("v0", "v1", 0)     #e1
        graph.add_edge("v0", "v1", 1)     #e2
        graph.add_edge("v1", "v3", 2)     #e3
        graph.add_edge("v1", "v3", 3)     #e4
        graph.add_edge("v1", "v3", 4)     #e5
        graph.add_edge("v3", "v2", 5)     #e6
        graph.add_edge("v2", "v0", 6)     #e7
        graph.add_edge("v2", "v0", 7)     #e8
        graph.add_edge("v2", "v0", 8)     #e9
        graph.add_edge("v1", "v5", 9)     #e10
        graph.add_edge("v1", "v5", 10)    #e11
        graph.add_edge("v1", "v5", 11)    #e12
        graph.add_edge("v5", "v4", 12)    #e13
        graph.add_edge("v0", "v4", 13)    #e14
        graph.add_edge("v0", "v4", 14)    #e15
        graph.add_edge("v0", "v4", 15)    #e16

        ne = len(graph.edges)
        e = np.identity(ne)
        alpha = np.array([
            e[0] - e[1],                #alpha1
            e[6] - e[7],                #alpha2
            e[3] - e[4],                #alpha3
            e[14]- e[15]+ e[6] - e[7],  #alpha4
            e[9] - e[11]+ e[3] - e[4],  #alpha5
            e[0] + e[5] + e[2] + e[6],  #alpha6
            e[0] + e[5] + e[3] + e[8],  #alpha7
            e[0] + e[5] + e[4] + e[7],  #alpha8
            e[1] + e[12]+ e[9] + e[13], #alpha9
            e[1] + e[12]+ e[10]+ e[14], #alpha10
            e[1] + e[12]+ e[11]+ e[15]  #alpha11
        ])
    elif struct_type == "bcc":
        graph.add_node("v0")
        graph.add_node("v1")
        graph.add_node("v2")
        graph.add_node("v3")
        graph.add_node("v4")
        graph.add_node("v5")
        graph.add_edge("v0", "v3", 0)     #e1
        graph.add_edge("v3", "v4", 1)     #e2
        graph.add_edge("v4", "v0", 2)     #e3
        graph.add_edge("v5", "v2", 3)     #e4
        graph.add_edge("v2", "v1", 4)     #e5
        graph.add_edge("v1", "v5", 5)     #e6
        graph.add_edge("v0", "v2", 6)     #e7
        graph.add_edge("v2", "v4", 7)     #e8
        graph.add_edge("v4", "v5", 8)     #e9
        graph.add_edge("v5", "v3", 9)     #e10
        graph.add_edge("v3", "v1", 10)    #e11
        graph.add_edge("v1", "v0", 11)    #e12
        
        ne = len(graph.edges)
        e = np.identity(ne)
        alpha = np.array([
            e[0] + e[10]+ e[11],                        #alpha1
            e[1] + e[8] + e[9],                         #alpha2
            e[2] + e[6] + e[7],                         #alpha3
            e[4] - e[7] + e[1] - e[10],                 #alpha4
            e[5] - e[8] + e[2] - e[11],                 #alpha5
            e[6] - e[3] - e[0] + e[9],                  #alpha6
            e[6] + e[7] + e[8] + e[9] + e[10] + e[11]   #alpha7
        ])
    else:
        raise Exception('Error:Please select "fcc", "hcp", or "bcc" for struct_type')
    tree = nx.maximum_spanning_tree(nx.Graph(graph))
    betti = ne - len(tree.edges(data=True))

    return [graph, alpha, e]

def calc_standard_realization(alpha, e, dim=3):
    matrixG0 = alpha @ alpha.T
    matrixb = alpha @ e
    matrixa = np.linalg.inv(matrixG0) @ matrixb
    matrixG11 = matrixG0[0 : dim, 0 :dim]
    matrixG22 = matrixG0[dim :, dim :]
    matrixG12 = matrixG0[0 : dim, dim : ]
    matrixG21 = matrixG0[dim :, 0 : dim]
    matrixG = matrixG11 - matrixG12 @ np.linalg.inv(matrixG22) @ matrixG21
    matrixPa = matrixa[ : dim, ]
    p = np.linalg.cholesky(matrixG)
    ed = ((matrixPa).T @ p)
    
    return [matrixG0, matrixG, matrixPa, ed, p]

def calc_coords(graph, ed):
    coords = [np.array([0, 0, 0])]
    nodes = list(graph.nodes)
    edge_keys = {edge[:2]:edge[2] for edge in graph.edges}
    for start_node, end_node, edge_index in graph.edges:
        path = nx.shortest_path(graph, source=nodes[0], target=start_node)        
        coord_displacement = np.sum([ed[edge_keys[path[i], path[i+1]]] 
                                    for i in range(len(path) - 1)], axis=0)    
        coord = coord_displacement + ed[edge_index]
        coords.append(coord)
    
    return np.array(coords)

def gen_structure(p, coords, volume=10, element_symbol="H"):
    elems = [Element(element_symbol)] * len(coords)
    struct = Structure(Lattice(p), elems, coords, coords_are_cartesian=True)
    struct.merge_sites(mode='delete', tol=0.01)
    struct.scale_lattice(volume)
    
    return struct


struct_type = "bcc"

graph, alpha, e = set_graph(struct_type)

matrixG0, matrixG, matrixPa, ed, p = calc_standard_realization(alpha, e)

coords = calc_coords(graph, ed)
dual_struct = gen_structure(p, coords)
dual_struct.to("dual_{}.cif".format(struct_type), symprec=0.1)
\end{lstlisting}